\documentclass[prd,aps,a4paper,nofootinbib,eqsecnum,twocolumn,floatfix]{revtex4}  

\newif\ifusesec
\usesectrue  
   
\usepackage{graphicx} 
\usepackage{mathrsfs}
\usepackage{amsmath,amsfonts,amssymb}
\usepackage{multirow}

\newcommand{\beq}{\begin{equation}}
\newcommand{\eeq}{\end{equation}}
\newcommand{\bea}{\begin{eqnarray}}
\newcommand{\eea}{\end{eqnarray}}

\begin{document}

\title{Projective path to points at infinity in spherically symmetric spacetimes}

\author{Donato \surname{Bini}$^{1}$, Giampiero Esposito$^{2,3}$}

\affiliation{$^1$Istituto per le Applicazioni del Calcolo ``M. Picone'', CNR, I-00185 Rome, Italy\\
Orcid: 0000-0002-5237-769X}
\affiliation{$^2$ Dipartimento di Fisica ``Ettore Pancini'', \\
Complesso Universitario di Monte S. Angelo,
Via Cintia Edificio 6, 80126 Napoli, Italy}
\affiliation{$^3$ Istituto Nazionale di Fisica Nucleare, Sezione di Napoli,\\
Complesso Universitario di Monte S. Angelo,
Via Cintia Edificio 6, 80126 Napoli, Italy \\
Orcid: 0000-0001-5930-8366} 

\begin{abstract}
This paper proves that, in a four-dimensional
spherically symmetric spacetime manifold,
one can consider coordinate transformations expressed
by fractional linear maps which give rise to isometries
and are the simplest example of coordinate transformation used
to bring infinity down to a finite distance.
The projective boundary of spherically symmetric spacetimes here studied is the
disjoint union of three points:
future timelike infinity, past timelike infinity, spacelike infinity, 
and the three-dimensional products of half-lines with a $2$-sphere.  
Geodesics are then studied in the 
projectively transformed $(t',r',\theta',\phi')$ coordinates
for Schwarzschild spacetime, with special interest in their 
way of approaching our points at infinity. 
Next, Nariai, de Sitter and G\"{o}del 
spacetimes are studied with our projective method. 
Since the kinds of infinity here defined depend only on the symmetry
of interest in a spacetime manifold, they have a broad range of
applications, which motivate the innovative analysis of Schwarzschild,
Nariai, de Sitter and G\"{o}del spacetimes.
\end{abstract}

\date{\today}

\maketitle

\section{Introduction}

The concept of points at infinity comes into play when studying 
several key features of theoretical physics:
\vskip 0.3cm
\noindent
(i) When one assumes that every isolated dynamical system is
described by the action principle, integration by parts in 
the action functional leads to the integral of the divergence
of a vector field. In turn, the divergence theorem leads 
to a surface integral where the surface is often regarded
as the surface at infinity (unless one deals with finite
regions). 
\vskip 0.3cm
\noindent
(ii) A classical spacetime can be said to be singularity-free
\cite{HE} if its timelike and null geodesics can be extended to arbitrary
values of the affine parameter. One therefore expects that, in
such a case, freely falling observers and photons should reach
the points at infinity.
\vskip 0.3cm
\noindent
(iii) Isolated gravitating systems are meant to be sufficiently
far from any interaction region, hence one is (implicitly) assuming
that the notion of points at infinity has been defined.
\vskip 0.3cm
\noindent
(iv) The coordinate-free definition of asymptotic flatness~\cite{HE}.
\vskip 0.3cm
\noindent
(v) Positive mass theorems~\cite{Chrusciel}.
\vskip 0.3cm
\noindent
(vi) In the global approach to quantum field theory~\cite{DeWitt}, the 
full specification of the space of field
histories is only achieved if suitable fall-off conditions
at infinity are assumed.  

In order to be more concrete, we would like to understand
how to describe the points at infinity, and 
how to derive from first principles why
a part of infinity consists of isolated points, whereas another
part of infinity consists of a three-dimensional manifold. In
this paper the methods of projective geometry are applied in
order to bring infinity down to a finite distance, and hence
answer the above questions. Our geometric definition of infinity
will be applied to spacetime models with spherical symmetry, 
with a final hint to
the G\"{o}del universe, which is instead cylindrically symmetric.
Interestingly, we will be able
to define points at infinity even when the more familiar 
construction of conformal boundary does not exist
(see our analysis of Nariai spacetime in Sec. V). Thus, our
framework might have valuable applications
to the asymptotic structure of spacetime in classical
and quantum gravity.
After this physics-oriented (and motivational) outline, we can present
our analysis in a systematic way, as follows.

The appropriate definition of points at infinity has always attracted
the attention of analysts and geometers in pure mathematics on the
one hand, and general relativists on the other hand. The latter
community is by now familiar with the conformal treatment of infinity
conceived by Penrose, since his famous Les Houches lectures 
in the early sixties \cite{Penrose}. The associated Carter-Penrose diagrams
are a powerful tool for modern research in classical and quantum 
gravity, but it remains legitimate to consider other mathematical 
tools, not only for intellectual curiosity, but also because
it remains an open problem what sort of boundary should be attached 
to a generic spacetime manifold \cite{Eardley}. 

In our paper we focus on the possible application of well-established
concepts in projective geometry. Let us therefore consider for a moment
a problem in complex analysis, i.e., how to bring the point at infinity
to a finite distance by means of a suitable transformation. In the
case of two complex variables $z$ and $w$, the substitutions
\begin{equation}
z'={1 \over z},\qquad w'=w,
\label{(1.1)}
\end{equation}
\begin{equation}
z'=z,\qquad w'={1 \over w},
\label{(1.2)}
\end{equation}
\begin{equation}
z'={1 \over z},\qquad  w'={1 \over w},
\label{(1.3)}
\end{equation}
are, apparently, the natural extension of the transformation 
$z'={1 \over z}$ used successfully in the case of one complex 
variable. Indeed, when dealing with one variable only, the
substitution $z'={1 \over z}$ does not lead to inconsistencies,
because it is a homography, which is bijective without exceptions
(cf. Appendix A),
among the neighbourhoods of the homologous points $z=\infty$
and $z'=0$. By contrast, in the case of two (or even more)
variables, the transformations (1.1)-(1.3) are not homographies
(see below), hence they are not bijective without exceptions. The inconvenience
lies precisely in the fact that the one-to-one nature of the
correspondence no longer holds at the points at infinity \cite{Severi}.
It is therefore clear that, in order to bring infinity down to a
finite distance in the case of two complex variables, the most
convenient transformations are the homographies. They are not
only the simplest, but also the unique transformations which are
bijective without exceptions. They can be written in the 
fractional linear form
\begin{eqnarray}
z'&=&\frac{(M_{\; 0}^{1}+M_{\; 1}^{1}z+M_{\; 2}^{1}w)}
{(M_{\; 0}^{0}+M_{\; 1}^{0}z+M_{\; 2}^{0}w)}, 
\nonumber\\
w'&=&\frac{(M_{\; 0}^{2}+M_{\; 1}^{2}z+M_{\; 2}^{2}w)}
{(M_{\; 0}^{0}+M_{\; 1}^{0}z+M_{\; 2}^{0}w)} ,
\label{(1.4)}
\end{eqnarray}
where the matrix $M$ of coefficients belongs to the general linear
group ${\rm GL}(3,{\mathbb C})$ (this ensures the bijective
nature of the transformation). One can check directly that the
analytic transformations \eqref{(1.1)}-\eqref{(1.3)} 
are not of the projective type \eqref{(1.4)} (for example,
the values of $M_{\; 0}^{0}$, $M_{\; 1}^{0}$, $M_{\; 2}^{0}$
for which the denominator equals $z$ differ from the values
for which the denominator equals $w$).
When the infinity is made equivalent to points
at finite distance by means of homographies, one says that the
projective point of view has been adopted (the relevance of
this property for spacetime geometry will become clear in
the following sections).

Section II proves the immersion of spherically symmetric four-dimensional
spacetime into real projective space ${\mathbb R}P_{4}$,
and obtains three concepts of infinity in a fully projective way
for all spherically symmetric spacetimes. 
Section III applies such a construction to two-dimensional Schwarzschild 
spacetime. Section IV studies geodesics of Schwarzschild in those
coordinates for which the infinity has been 
brought to finite distance. 
The Nariai spacetime model is studied in Sec. V,
de Sitter metric is investigated in Sec. VI,
while Sec. VII considers briefly the G\"{o}del universe.
Open problems are discussed in Sec. VIII. 

\section{From projective coordinates to three concepts of infinity
in spherically symmetric spacetimes}

\subsection{Immersion in real projective space}

Let us begin by recalling that real projective space
${\mathbb R}P_{4}$ can be obtained from the 4-sphere $S^{4}$
by identifying diametrically opposite points. Thus, given the
five homogeneous coordinates $y^{0},...,y^{4}$ of
${\mathbb R}P_{4}$, they must obey the 4-sphere equation
\begin{equation}
\sum_{\lambda=0}^{4}(y^{\lambda})^{2}=1,
\label{(2.1)}
\end{equation}
jointly with the identification rule
\begin{equation}
y^{\lambda} \sim - y^{\lambda}, \qquad
\forall \lambda=0,1,2,3,4.
\label{(2.2)}
\end{equation}
Hence we consider an immersion\footnote{A further motivation
lies in the fact that real projective space can be
endowed with a spherically symmetric metric
\cite{V1960,V1963}.} 
of four-dimensional spherically symmetric
spacetime into ${\mathbb R}P_{4}$ defined by the equations
\begin{equation}
y^{\mu}=x^{\mu}y^{0}, \qquad \forall \mu=1,2,3,4,
\label{(2.3)}
\end{equation}
completed by the equation consisting of the 4-sphere condition
re-expressed therefore in the form 
\begin{equation}
y^{0}={1 \over \sqrt{1+\sum_{\mu=1}^{4}(x^{\mu})^{2}}}.
\label{(2.4)}
\end{equation}
Equations \eqref{(2.3)}-\eqref{(2.4)} define indeed an 
immersion, not to be confused with an embedding, 
because the associated Jacobian matrix has the form
\begin{equation}
{\partial y^{\mu}\over \partial x_{\nu}}=\delta^{\mu \nu} \; y^{0}
-x^{\mu}x^{\nu}(y^{0})^{3}, \; \forall \mu,\nu=1,...,4,
\label{(2.5)}
\end{equation}
\begin{equation}
{\partial y^{0}\over \partial x_{\nu}}=-x^{\nu}(y^{0})^{3}.
\label{(2.6)}
\end{equation}
Thus, at $x^{\mu}=0$ the Jacobian matrix reduces to
\begin{equation}
\left . {\partial y^{\mu}\over \partial x_{\nu}}
\right |_{0}={\rm diag}(1,1,1,1),
\label{(2.7)}
\end{equation}
\begin{equation}
\left . {\partial y^{0}\over \partial x_{\nu}}
\right |_{0}=0,
\label{(2.8)}
\end{equation}
and we find that such a matrix has rank $4$ at $x^\mu=0$,
because the sub-matrix \eqref{(2.7)} has non-vanishing determinant.
The immersion criterion in Sec. 1.8 of Ref. \cite{Nacinovich}
is therefore satisfied (see also Sec. 5.2.7 of Ref. \cite{Nakahara2003}).

In other words, we can legitimately assume that the coordinates
$x^\alpha=\{x^{1},x^{2},x^{3},x^{4}\}$ (with $\alpha=1,\ldots, 4$) 
used for a real four-dimensional spherically symmetric
spacetime manifold arise from five homogeneous coordinates
$y^A=\{y^{0},y^{1},y^{2},y^{3},y^{4}\}$ (with $A=0,\ldots, 4$) 
according to the defining relation 
(our $x^{1}$ will be the time coordinate, see below in 
Eq. \eqref{(2.13)}) $x^\alpha=\frac{y^\alpha}{y^0}$, 
that is\footnote{As a rule we assume $\alpha=1,\ldots 4$, and 
$A=0\,,\ldots 4$. Therefore $A=(0,\alpha)$ will also be 
a used notation. When working instead in a 
2-dimensional manifold we will use the notation $x^a=(x^1,x^2)$, i.e., $a=1,2$.}
\begin{equation}
x^{1}={y^{1}\over y^{0}}, \qquad
x^{2}={y^{2}\over y^{0}}, \qquad
x^{3}={y^{3}\over y^{0}}, \qquad
x^{4}={y^{4}\over y^{0}},
\label{(2.9)}
\end{equation}
the $y$'s being subject to the familiar linear transformations in ${\mathbb R}P_{4}$
(with constant coefficients)
\begin{equation}
{y'}^{B}=A^B{}_C y^C,\qquad {\rm det}\left(A_{\; C}^{B}\right) \not =0,
\label{(2.10)}
\end{equation}
which therefore imply the following transformation rules for spacetime
coordinates (cf. Eq. \eqref{(1.4)}):
\begin{equation}
{x'}^{\mu}=\frac{A^\mu{}_B y^B}{A^0{}_Cy^C},
 \qquad
\mu=1,\ldots, 4\,,\quad B,C=0,\ldots 4.
\label{(2.11)}
\end{equation}
Since only the ratios of $y^{B}$ coordinates are relevant,
we assume the standard identification
$$
y^{B} \sim \zeta y^{B}, \; \forall \zeta \in
{\mathbb R} - \{ 0 \}.
$$
In other words, we consider the linear transformations \eqref{(2.10)}
of homogeneous coordinates,
which induce in turn a particular set of coordinate transformations
in the spacetime manifold. 
Equations \eqref{(2.11)} provide a projective way of studying
points at infinity of a given manifold, as we are now going to show.
Our spacetime coordinate transformations \eqref{(2.11)} are
inspired by projective geometry but {\it do not describe an isometric
embedding of spherically symmetric spacetime into five- or higher-dimensional
manifolds} \cite{KasnerA,KasnerB,1971,2002,2003,2012}. 
However, we will prove in detail that
Eqs. \eqref{(2.11)} lead to isometries in four-dimensional
Schwarzschild spacetime, and this property is sufficient for
our purposes. 

Our approach to projective structures on curved manifolds
differs from the one based on the connection
\cite{Schouten,Tashiro}. 
An alternative to the linear transformation law \eqref{(2.10)}
might be provided by the use of homogeneous polynomials 
of a given degree on the right-hand side, for which
we refer the reader to Appendix B.  

For this purpose, let us focus on the cases in which, {\it in dimensionless
units}, local coordinates $t,r,\theta,\phi$ exist for which 
(these are the spherical coordinates, and for an enlightening
physical discussion we refer the reader to the work in
Ref. \cite{Bondi})
\begin{eqnarray}
\; & \; &
t \in {\mathbb R} \cup \left \{ -\infty, \infty \right \}, \; 
r \in {\mathbb R}^{+} \cup \left \{ \infty \right \}, 
\nonumber \\
& \; & \theta \in [0,\pi] , \;
\phi \in [0,2\pi[.
\label{(2.12)}
\end{eqnarray}
We limit ourselves to assuming, for the time being, the
existence of the coordinate ranges in \eqref{(2.12)}
(in other words, we arrive at a metric-independent formulation,
but we need suitable local coordinates according to Eq.
\eqref{(2.12)}).

Let us now introduce the notation
\begin{equation}
x^{1}=t, \; x^{2}=r, \; x^{3}=\theta, \; x^{4}=\phi.
\label{(2.13)}
\end{equation}
Hence Eq. \eqref{(2.11)} reads as (upon dividing numerator and
denominator by $y^{0}$)
\begin{equation}
{x'}^{\mu}={\Bigr(A_{\; 0}^{\mu}+A_{\; 1}^{\mu}t+A_{\; 2}^{\mu}r
+A_{\; 3}^{\mu}\theta+A_{\; 4}^{\mu}\phi \Bigr) \over
\Bigr(A_{\; 0}^{0}+A_{\; 1}^{0}t+A_{\; 2}^{0}r
+A_{\; 3}^{0}\theta+A_{\; 4}^{0}\phi \Bigr)}.
\label{(2.14)}
\end{equation}
Interestingly, at this stage the unphysical
homogeneous coordinates $y^{A}$ no longer occur, and we are dealing with 
a fractional linear transformation of spacetime coordinates.
For convenience, as stated above, we avoid 
here heavier notations and work with dimensionless quantities 
(indeed, dimensionless coordinates imply dimensionless coefficients $A_{\; \alpha}^{\mu}$).

An important remark is now in order. If two functions of real variables
are differentiable, then so is their quotient so long as the denominator
is not zero. Since polynomials are differentiable, therefore rational
functions of real variables will be differentiable whenever they are
defined. Thus, in order to see if a rational function is differentiable,
one has to check that the denominator is nowhere vanishing. However,
bearing in mind that, in one real variable, the fractional linear map
$$
h(x)=\frac{(ax+b)}{(cx+d)}
$$
can be defined also when the denominator vanishes by requiring that
$$
h \left(x=-\frac{d}{c}\right)=\infty,
$$
we can similarly define the coordinate transformation \eqref{(2.14)}
and its derivatives of arbitrary order for all values of 
$t,r,\theta,\phi$ upon requiring that, when 
$$
A_{\; 0}^{0}+A_{\; 1}^{0}t+A_{\; 2}^{0}r+A_{\; 3}^{0}\theta
+A_{\; 4}^{0}\phi=0,
$$
the coordinate ${x'}^{\mu}$ is infinite.

Last, let us note that the coordinate transformation \eqref{(2.14)}, despite 
being mathematically allowed, should in general satisfy additional requirements 
to be also physically admissible \cite{Bini:2012ht}.  For example (and for 
completeness as well), we mention that one has to check that the new 
metric component $g_{t't'}<0$ 
and that the matrix
$M_{i'j'}=g_{i'j'}-\frac{g_{t'i'}g_{t'j'}}{g_{t't'}}$ is positive definite, 
i.e., all principal minors have a positive determinant
\begin{equation}
M_{1'1'}>0,\qquad M_{1'1'}M_{2'2'}-(M_{1'2'})^2>0,
\end{equation}
and
\begin{equation}
{\rm det}[M_{i'j'}]>0.
\end{equation}
Fulfilling the above requirements (or equivalent requirements) is more 
conveniently done case by case, i.e., when working within a special starting 
metric. However, having always at disposal a sufficiently large number of 
parameters in the coordinate transformation \eqref{(2.14)}, the admissibility 
of the corresponding coordinates can be easily checked in the neighbourhood 
of any arbitrary point with a special choice of the parameters. We will not 
insist any longer then on this requirement.
  
\subsection{The ${\rm GL}(5,{\mathbb R})$ matrix $A$}

In light of Eq. \eqref{(2.14)} we realize that, since 
$\theta$ and $\phi$ are varying in finite-measure intervals,
they do not affect the values of the limits of
${x'}^{\mu}$ when $t \rightarrow \pm \infty$, or as
$r \rightarrow \infty$ ({\it here, such limits will be
considered one at a time, i.e., infinite $t$ at fixed $r$,
or infinite $r$ at fixed $t$, in order to deal with
meaningful expressions}). Hence we obtain the formulas 
\begin{eqnarray}
\lim_{t \to \pm \infty}t'=\lim_{t \to \pm \infty}
{A_{\; 1}^{1}t \over A_{\; 1}^{0}t}
={A_{\; 1}^{1}\over A_{\; 1}^{0}}, 
\nonumber \\
\lim_{r \to \infty}r'=\lim_{r \to \infty}
{A_{\; 2}^{2}r \over A_{\; 2}^{0}r}
={A_{\; 2}^{2} \over A_{\; 2}^{0}},
\label{(2.16)}
\end{eqnarray}
which show clearly that our points at timelike or spacelike
infinity depend only on four matrix elements of $A$.
At this stage we can therefore assume, without loss of
generality, the following form of the matrix $A$:
\begin{equation}
A =\left(\begin{matrix}
1 & A_{\; 1}^{0} & A_{\; 2}^{0} & 0 & 0 \cr
0 & 1 & 0 & 0 & 0 \cr
0 & 0 & 1 & 0 & 0 \cr
0 & 0 & 0 & 1 & 0 \cr
0 & 0 & 0 & 0 & 1
\end{matrix}\right),
\label{(2.17)}
\end{equation}
where the values of $A_{\; 1}^{0}$ and $A_{\; 2}^{0}$ do not
affect the ${\rm GL}(5,{\mathbb R})$ nature of $A$. The form of $A$
might be restricted upon requiring that the quintuplet
$y'=Ay$ in Eq. \eqref{(2.10)} also belongs to the 4-sphere, but
this would be incorrect because the projective transformations
of homogeneous coordinates are not norm-preserving.

To sum up, anyone willing to use our method can exploit
the two-parameter family \eqref{(2.17)} of matrices of ${\rm GL}(5,{\mathbb R})$
in all subsequent calculations with the understanding that,
when the behaviour of geodesics is studied, the $(r,t)$ values
for which $1+A_{\; 1}^{0}t+A_{\; 2}^{0}r$ vanishes 
correspond to singularities of the coordinate transformation.
These singularities are harmless in light of the remarks at the
end of subsection IIA, and are lines in the $(t,r)$ plane.

\subsection{Three kinds of projective infinity}

Another merit of this choice of $A$ is its manifest link
with the choice of advanced ($t+r$) 
or retarded ($t-r$) coordinates,
depending on whether the free parameters 
$A_{\; 1}^{0}$ and $A_{\; 2}^{0}$
are equal or opposite to each other.
Such a matrix remains invertible for all values of $A_{\; 1}^{0}$
and $A_{\; 2}^{0}$, since its determinant equals $1$.
Hence we obtain in turn
\begin{equation}
{t' \over t}={r' \over r}={\theta' \over \theta}
={\phi' \over \phi}=\Bigr(1+A_{\; 1}^{0}t+A_{\; 2}^{0}r
\Bigr)^{-1},
\label{(2.18)}
\end{equation}
jointly with the limits
\begin{equation}
\lim_{t \to \pm \infty}t'
={t'}_{\infty}, \qquad \lim_{r \to \infty}t'=0, 
\label{(2.19)}
\end{equation}
\begin{equation}
\lim_{r \to \infty}r'
={r'}_{\infty},
\qquad \lim_{t \to \pm \infty}r'=0,
\label{(2.20)}
\end{equation}
\begin{equation}
\lim_{t \to \pm \infty}\theta'=0=\lim_{r \to \infty} \theta' ,
\label{(2.21)}
\end{equation}
\begin{equation}
\lim_{t \to \pm \infty}\phi'=0=\lim_{r \to \infty} \phi' \,,
\label{(2.22)}
\end{equation}
where we have defined 
\begin{equation}
{t'}_{\infty}={1 \over A_{\; 1}^{0}},\qquad
{r'}_{\infty}={1 \over A_{\; 2}^{0}}.
\label{(2.23)}
\end{equation}

We can therefore define the first two concepts of infinity
(cf. Ref. \cite{1974}):
\vskip 0.3cm
\noindent
(1) Projective timelike infinity, i.e., the point having coordinates
\begin{equation}
\left(t'={1 \over A_{\; 1}^{0}}={t'}_{\infty},r'=0,\theta'=0,\phi'=0 \right).
\label{(2.24)}
\end{equation}
According to the sign of $A_{\; 1}^{0}$, we can further distinguish 
among future and past timelike infinity.
\vskip 0.3cm
\noindent
(2) Projective spacelike infinity, which consists of the point with coordinates
\begin{equation}
\left(t'=0,r'={1 \over A_{\; 2}^{0}}={r}'_{\infty},\theta'=0,\phi'=0 \right).
\label{(2.25)}
\end{equation}
As we said before, singularities of our
coordinate transformation correspond to the zeros
of the denominator, i.e., the points $(r,t)$ for which
\begin{equation}
1+A_{\; 1}^{0}t+A_{\; 2}^{0}r=0.
\label{(2.26)}
\end{equation}
Such an equation is solved by
\begin{equation}
t=-{t}'_{\infty} \left(1+{r \over {r}'_{\infty}}\right).
\label{(2.27)}
\end{equation}

Furthermore, it is also of physical interest to study the case in which
the projective map preserves the $2$-sphere, where
$\theta$ and $\phi$ are angular coordinates. We then find
\begin{eqnarray}
\; & \; &
\theta'=\theta, \; \phi'=\phi \Longrightarrow
A_{\; 1}^{0}t+A_{\; 2}^{0}r=0 
\nonumber \\
& \; & \Longrightarrow
t=-{A_{\; 2}^{0}\over A_{\; 1}^{0}}r=-{t}'_{\infty}
{r \over {r}'_{\infty}}.
\label{(2.28)}
\end{eqnarray}
At this stage, we deal with
the projective counterpart of null infinity. Indeed,
when Eq. \eqref{(2.28)} holds,
Eq. \eqref{(2.18)} also implies that $t'=t$, $r'=r$, and hence the coordinates
of points at infinity are of the third kind: 
\vskip 0.3cm
\noindent
(3) Product of a half-line departing from the origin in the
first or fourth quadrant, with a $2$-sphere:
\begin{equation}
\left(t'=t=-{t}'_{\infty}{r \over {r}'_{\infty}},r'=r,\theta'=\theta,\phi'=\phi \right).
\label{(2.29)}
\end{equation}
In other words, we can say that this kind of infinity has points belonging
to the products $\rho_{1}\times S^{2}$ and $\rho_{2} \times S^{2}$,
where, in the plane with abscissa $r'$ and ordinate $t'$, the negative value of
${t}'_{\infty}$ generates the half-line $\rho_{1}$ which departs from the origin
and lies in the first quadrant, whereas the positive value of
${t}'_{\infty}$ generates the half-line $\rho_{2}$ which starts from 
the origin and lies in the fourth quadrant. Our original construction 
(3) is appropriate for Schwarzschild geometry and is  
the projective counterpart of future and past null infinity in
general relativity. Upon defining 
\begin{eqnarray*}
U^{1}&=& t,\nonumber\\
U^{2}&=&r,\nonumber\\
U^{3}&=&\sin(\theta)\cos(\phi),\nonumber\\
U^{4}&=&\sin(\theta)\sin(\phi),\nonumber\\
U^{5}&=&\cos(\theta),
\end{eqnarray*}
we are dealing with the three-dimensional submanifold of $R^5$ defined by
the equations
\begin{equation}
F(U^1,U^2)=A_{\; 1}^{0}U^{1}+A_{\; 2}^{0}U^{2}=0,
\label{(2.30)}
\end{equation}
\begin{equation}
H(U^3,U^4,U^5)=\sum_{k=3}^{5}(U^{k})^{2}-1=0.
\label{(2.31)}
\end{equation}
Such equations can be studied all at once by looking at the
level sets (here $\lambda$ and $\mu$ are real parameters)
$$
\lambda F(U^1,U^2)+\mu H(U^3,U^4,U^5)={\rm constant},
$$
where vanishing values of $\mu$ and $\lambda$ correspond to
Eqs. \eqref{(2.30)} and \eqref{(2.31)}, respectively.
At a deeper level, not exploited here but certainly original,
we are preparing the ground for studying null infinity from
the point of view of the rich mathematical theory of
algebraic surfaces \cite{Enriques,Zariski}.

Our projective definitions
of infinity remain valid whenever spherical coordinates can be
used. Important applications will be discussed in Secs. IV, 
V and VI. To sum up, the  projective  boundary of spherically 
symmetric spacetime models here studied is the
disjoint union of three points:
future timelike infinity, past timelike infinity, spacelike infinity, 
and the three-dimensional products $\rho_{1} \times S^{2}$ and
$\rho_{2} \times S^{2}$. 
Note also that, by comparison of 
\eqref{(2.26)} and \eqref{(2.27)}, 
if ${t}'_{\infty}$ is positive, the third kind of infinity 
is reached before reaching the zeros of the denominator.

In Secs. III, IV, V and VI, where $t,r,\theta,\phi$ will be also 
solutions of the geodesic equation for the chosen metric which
solves the Einstein equations, the resulting (numerical) analysis
will face the additional problem of the joint effect of $t$
and $r$ in reaching the line of singular points of 
Eq. \eqref{(2.26)}. 

\subsection{Observers at rest in spherical coordinates and 
in motion in projective coordinates}

Let us consider for simplicity the coordinate map \eqref{(2.9)} from spherical 
coordinates $(t,r,\theta,\phi)$ to projective coordinates $(t',r',\theta',\phi')$ 
\begin{eqnarray}
t'&=& \frac{t}{\rho}\,,\qquad r'=\frac{r}{\rho},\qquad
\theta'=\frac{\theta}{\rho}\,,\qquad \phi'=\frac{\phi}{\rho},\qquad
\label{(2.32)}
\end{eqnarray}
with $\rho=1+A_{\; 1}^{0}t+A_{\; 2}^{0}r=(1-A^0{}_1t'-A^0{}_2r')^{-1}$.
We find
\begin{equation}
\frac{\partial}{\partial t} 
=\frac{1}{\rho}\left(1- t'A^0{}_1\right) \frac{\partial}{\partial t'}
-\frac{A^0{}_1}{\rho}{\mathbf n}
\label{(2.33)}
\end{equation}
where we have defined
\begin{equation}
{\mathbf n}=r' \frac{\partial}{\partial r'}
+ \theta' \frac{\partial}{\partial \theta'}
+ \phi'\frac{\partial}{\partial \phi'}.
\label{(2.34)}
\end{equation}
Let us denote by $g_{\alpha\beta}$ the 
components of the spacetime metric referred to spherical 
coordinates and by $g_{\alpha'\beta'}$ the components of the 
transformed metric in projective coordinates.
The four-velocity of an observer at rest with respect to spherical coordinates is
\begin{equation}
u_{\rm sc}=\frac{1}{\sqrt{-g_{tt}}}\frac{\partial}{\partial t} ,
\label{(2.35)}
\end{equation}
whereas that of an observer at rest with respect to projective coordinates is
\begin{equation}
u_{\rm pc}=\frac{1}{\sqrt{-g_{t't'}}} \frac{\partial}{\partial t'} .
\label{(2.36)}
\end{equation}
Equation \eqref{(2.35)} can then be cast in the form
\begin{equation}
u_{\rm sc}
= \gamma(u_{\rm sc},u_{\rm pc})[u_{\rm pc}+\nu(u_{\rm sc},u_{\rm pc})],
\label{(2.37)}
\end{equation}
where the relative velocity $\nu(u_{\rm sc},u_{\rm pc})$ is orthogonal 
to $u_{\rm pc}$ by definition (as the notation suggests) and, in general, 
it is not aligned along the spatial vector ${\mathbf n}$ since the metric is nondiagonal.
Equation \eqref{(2.37)} represents the boost from $u_{\rm pc}$  to $u_{\rm sc}$. 
The analytical and geometrical properties of this boost offer 
another point of view on how to look at the coordinate map. 
We note that the expression \eqref{(2.37)} is completely general and hence 
rather involved. It implies that an observer at rest in the projective 
coordinates moves (in all spatial directions) with respect to one at rest 
in the original spherical coordinate set. Moreover, in order to remain 
at rest in the projective coordinate system such an observer needs 
accelerations, even if he/she were geodesic in the original coordinate set.
We will discuss this type of characteristics in 
Eq. \eqref{(3.18)} corresponding to the simplest 
situation of a two-dimensional Schwarzschild spacetime.
We will, however, always provide the new metric component $g_{t't'}$ 
needed in Eq. \eqref{(2.36)} in all the considered examples.

A complementary point of view consists in studying the foliation of spacetime  
by the hypersurfaces $t'$=constant (instead of the worldlines 
$t'=$ variable). In this case we find
\begin{equation}
dt'=\frac{(1+rA^0{}_2)dt+t A^0{}_2 dr}{\rho^2}.
\end{equation}
Slicing observers (\lq\lq sli") with four-velocity 1-form field parallel to $dt'$,
\begin{equation}
\label{eq_sli_obs}
u_{\rm sli}=-N' dt'=-\frac{N'}{\rho^2} [(1+rA^0{}_2)dt+t A^0{}_2 dr]\,,
\end{equation}
where $N'=1/\sqrt{-g^{t't'}}$ is a normalization factor ensuring $u_{\rm sli}\cdot 
u_{\rm sli}=-1$, are vorticity-free by definition, and $du_{\rm sli}=-N'_{,a'} dx^{a'}\wedge 
dt'\equiv - u_{\rm sli}\wedge a( u_{\rm sli})$. The latter relation identifies 
the acceleration of slicing observers as
\begin{equation}
a( u_{\rm sli})=(\partial_{a'}\ln N')dx^{a'}.
\end{equation}

\subsection{The standard concepts of infinity for Schwarzschild}

In order to help the general reader who might want to compare
our method with standard techniques 
we here recall that, upon defining the retarded coordinate
\begin{equation}
u=t-r_*,
\label{(2.38)}
\end{equation}
and the advanced coordinate
\begin{equation}
v=t+r_*,
\label{(2.39)}
\end{equation}
with
\begin{equation}
r_*=r \left[1+\frac{2M}{r}\log \left( \frac{r}{2M}-1 \right)\right]
\label{(2.40)}
\end{equation}
the so-called \lq\lq tortoise" coordinate,
one can say that future null infinity of Schwarzschild spacetime
has topology $S^{2} \times \mathbb{R}$ and
is defined by the conditions \cite{Schmidt}
\begin{equation}
v \rightarrow \infty, \; u \in ]-\infty,+\infty[,
\label{(2.41)}
\end{equation}
while for past null infinity (having the same topology)
one has, conversely,
\begin{equation}
u \rightarrow -\infty, \; v \in ]-\infty,+\infty[.
\label{(2.42)}
\end{equation}
By contrast, spacelike infinity is a point obeying the condition
\begin{equation}
v \rightarrow +\infty, \; u \rightarrow - \infty, \;
v+u={\rm finite}.
\label{(2.43)}
\end{equation}

The substantial difference between our projective construction of
infinity and the definition summarized by Eqs. 
\eqref{(2.39)}-\eqref{(2.43)} lies
in the fact that {\it we only need the assumption of spherical symmetry
in order to define it,
but we do not rely on solutions of the Einstein equations}. Our projective 
definition can be applied to any spherically symmetric spacetime.
In other words, {\it in our framework the metric which solves Einstein 
equations is used at a later stage, in order to study geodesics and
how they approach the points at infinity. Our projective method does
not need the tortoise coordinate \eqref{(2.40)}, and our null
infinity is not the boundary of the Penrose conformal completion of spacetime}
(see the following subsection).

\subsection{Penrose conformal completion of spacetime}

On a manifold $(M,g)$ of dimension $n$, a diffeomorphism $f: M \rightarrow M$
is an {\it isometry} if it preserves the metric. If $x^{\mu}$ 
are the coordinates of a point $p$ of $M$, and if $y^{\alpha}$
are the coordinates of $f(p)$, the expression in components 
reads as \cite{Nakahara2003}
\begin{equation}
{\partial y^{\alpha}\over \partial x^{\mu}}
{\partial y^{\beta}\over \partial x^{\nu}}g_{\alpha \beta}(f(p))
=g_{\mu \nu}(p),
\label{(2.44)}
\end{equation}
where all indices run from $0$ through $(n-1)$ 
if $g$ has Lorentzian signature,
which leads to the desired preservation condition
\begin{equation}
g_{\alpha \beta}(y(x))dy^{\alpha}dy^{\beta}
=g_{\mu \nu}(x)dx^{\mu}dx^{\nu}.
\label{(2.45)}
\end{equation}
The diffeomorphism $f:M \rightarrow M$ is instead a
{\it conformal transformation} 
\cite{Nakahara2003} if it preserves the metric up to
a scale factor, i.e.,
\begin{equation}
{\partial y^{\alpha}\over \partial x^{\mu}}
{\partial y^{\beta}\over \partial x^{\nu}}g_{\alpha \beta}(f(p))
=\Omega^{2}(p)g_{\mu \nu}(p),
\label{(2.46)}
\end{equation}
which leads in turn to
\begin{equation}
g_{\alpha \beta}(y(x))dy^{\alpha}dy^{\beta}=\Omega^{2}(x)g_{\mu \nu}(x)
dx^{\mu}dx^{\nu}.
\label{(2.47)}
\end{equation}
Note that, so far, only one metric, $g$, is involved in the definitions, 
but in different coordinates.

Given instead two metrics $g$ and $h$ on $M$, one says that
$h$ is {\it conformally related} to $g$ if
\begin{equation}
h_{\mu \nu}(p)=\Omega^{2}(p)g_{\mu \nu}(p).
\label{(2.48)}
\end{equation}
The transformation \eqref{(2.48)} is a {\it conformal rescaling},
or {\it Weyl rescaling}. In the literature, unfortunately, the
distinction between conformal transformations and conformal rescalings
is not always clearly presented.

The Penrose method considers a physical spacetime 
$({\hat M},{\hat g})$, a manifold $M$ with topological boundary
${\cal I}={\partial M}$, a metric $g$ for $M$, a diffeomorphism from
${\hat M}$ to $M \setminus {\cal I}$, by means of which ${\hat M}$ is
identified with $M \setminus {\cal I}$
\cite{Ashtekar2014}. It is then possible to consider
a smooth function $\Omega$ on $M$, with
\begin{equation}
g=\Omega^{2}{\hat g} \; {\rm on} \; {\hat M}=M \setminus {\cal I},
\label{(2.49)}
\end{equation}
and the function $\Omega$ should behave as follows:
\begin{equation}
\Omega=0 \; {\rm on} \; {\cal I}={\partial M}, \;
\nabla_{\mu}\Omega \not = 0 \; {\rm on} \; 
{\cal I}={\partial M}.
\label{(2.50)}
\end{equation}
The manifold ${\cal I}={\partial M}$ must have topology
$S^{2} \times {\mathbb R}$, while the metric $\hat g$ solves
the Einstein equations on $\hat M$. 

To sum up, Penrose relies upon conformal rescalings which
relate the metric on $\hat M$ to the metric on $M$. {\it We build
instead isometries on physical spacetime. Such isometries can
be viewed as conformal transformations with $\Omega=1$ on
physical spacetime, but have empty intersection with the set
of conformal rescalings considered by Penrose.}

\begin{figure}
\includegraphics[scale=0.3]{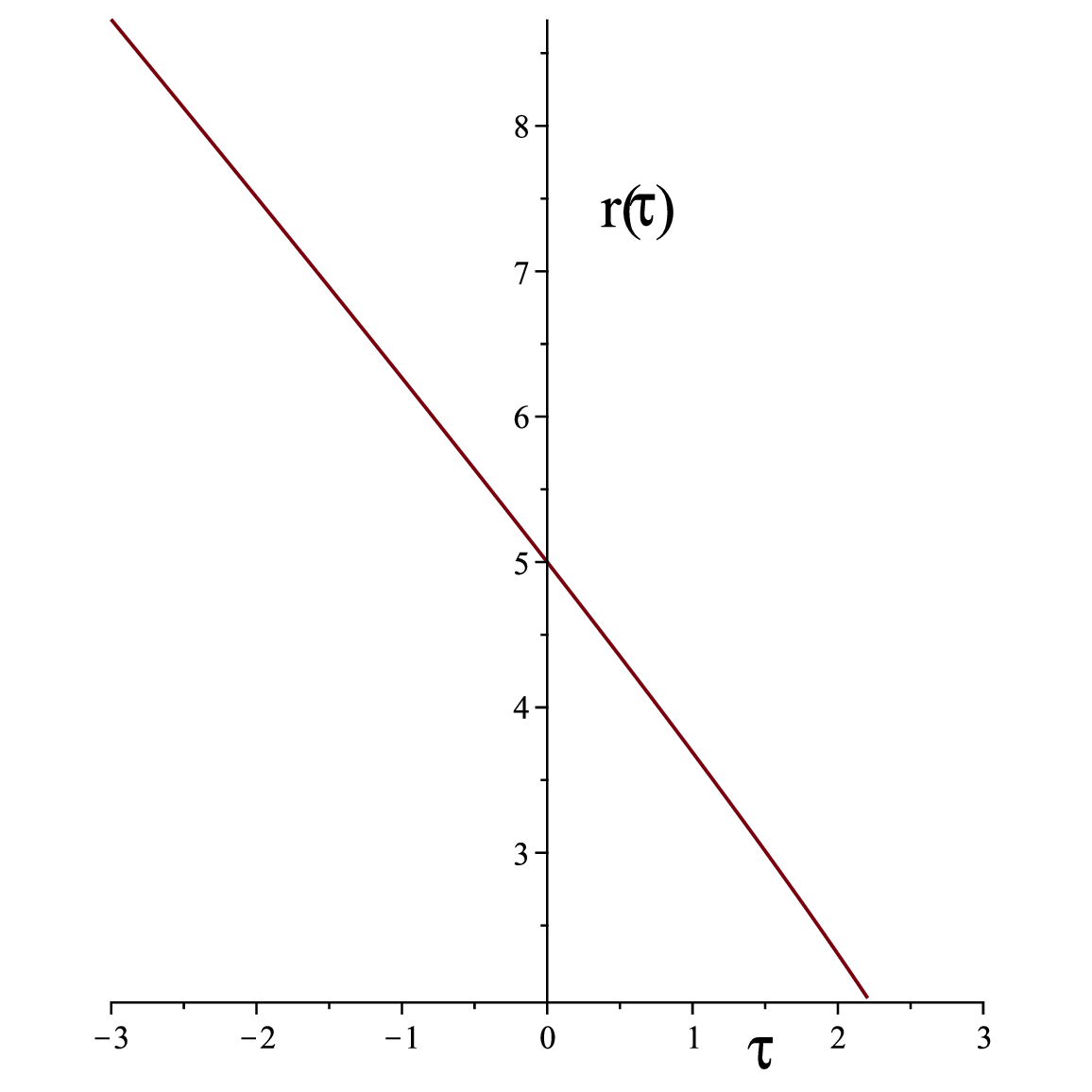}
\includegraphics[scale=0.3]{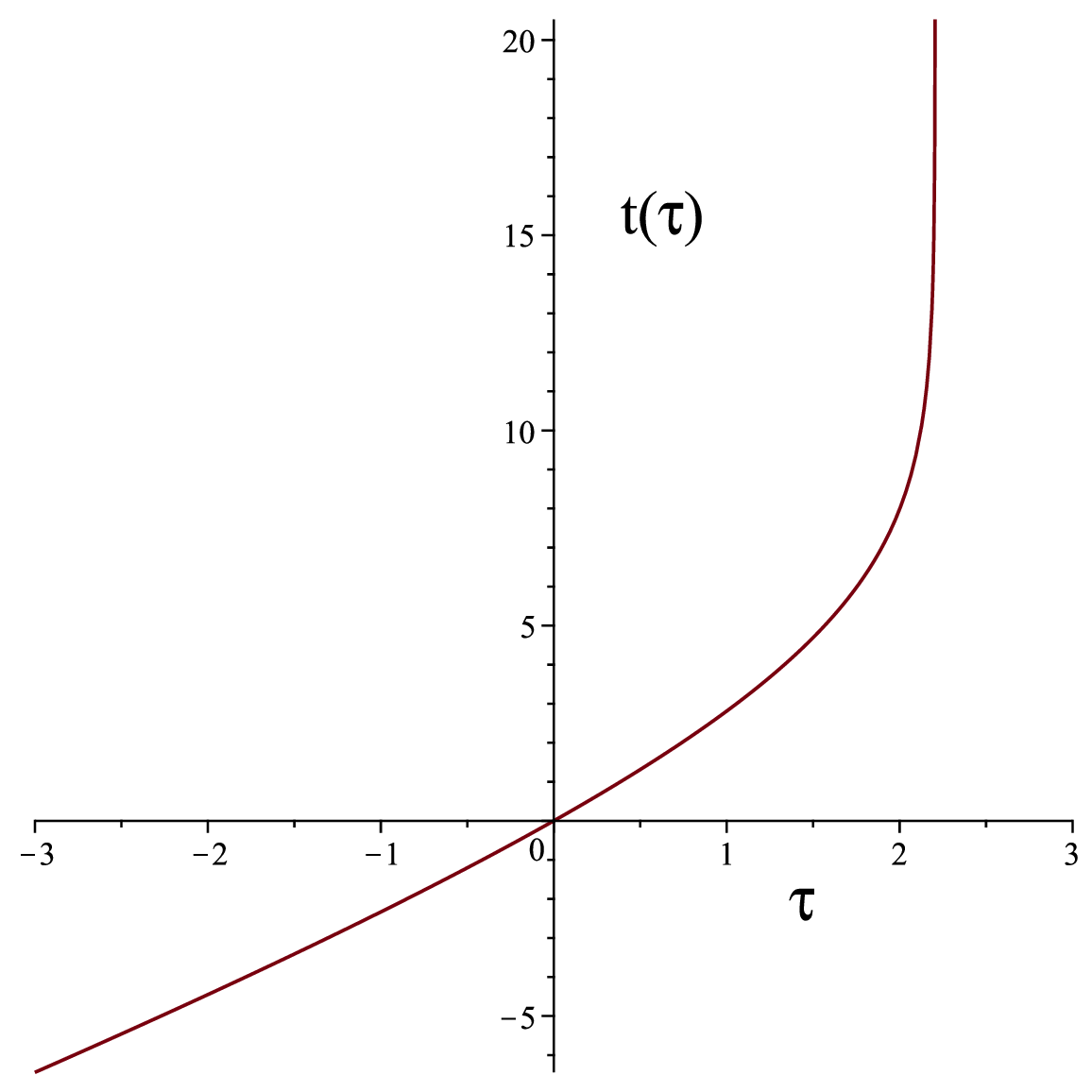}
\caption{\label{fig:1} Numerical integration of the timelike  
($\epsilon=1$ in Eq. \eqref{geo_eqs_sch2d}) geodesics in a 2-dimensional 
Schwarzschild spacetime, with parameters $M=1,E=3/2$ and 
initial conditions $r(0)=5,t(0)=0$. 
One sees (numerically) that when $\tau_*=2.2053644$ the particle 
has reached $r=2$ while $t(\tau_*)\approx 38.3832$.}
\end{figure}

\begin{figure}
\includegraphics[scale=0.3]{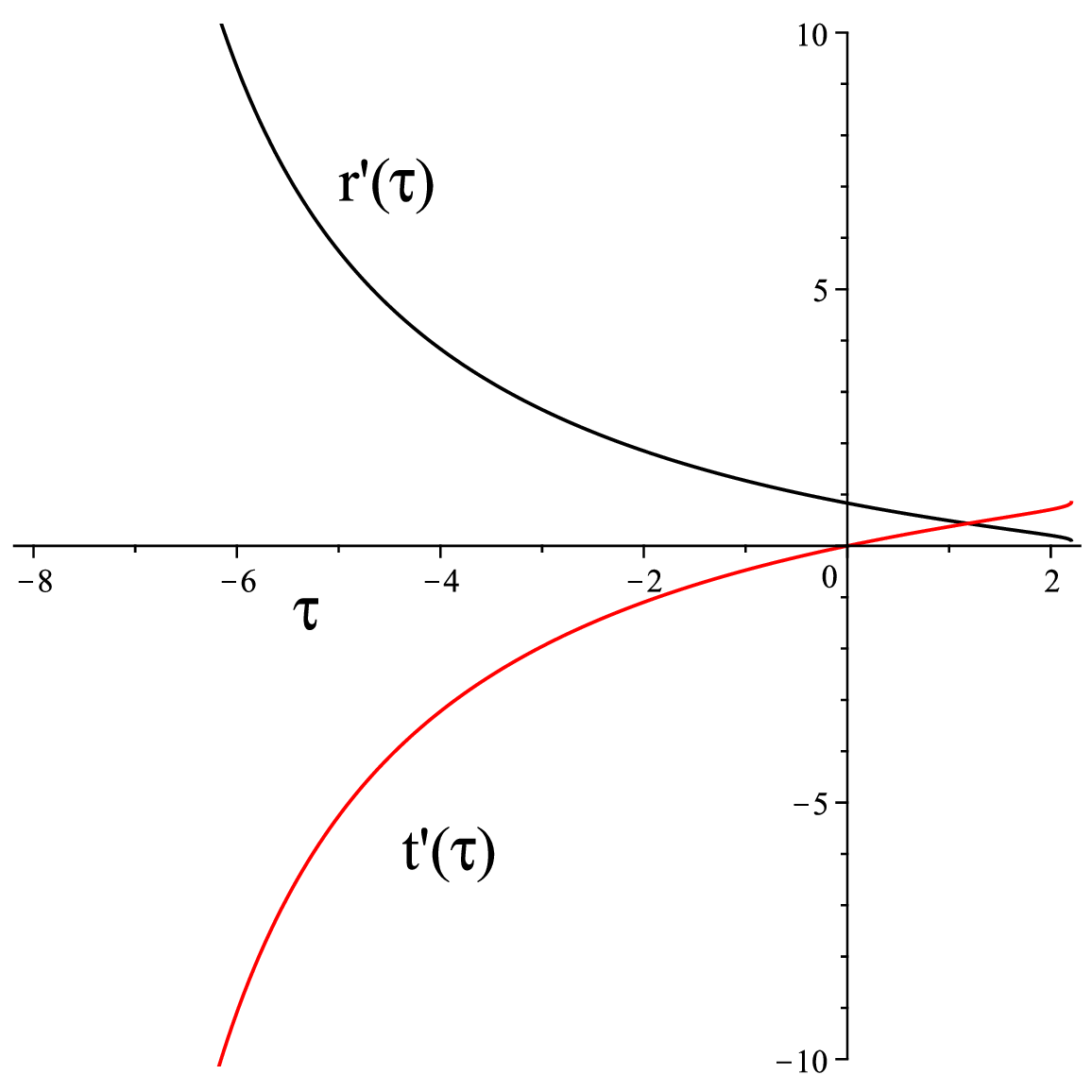}
\includegraphics[scale=0.3]{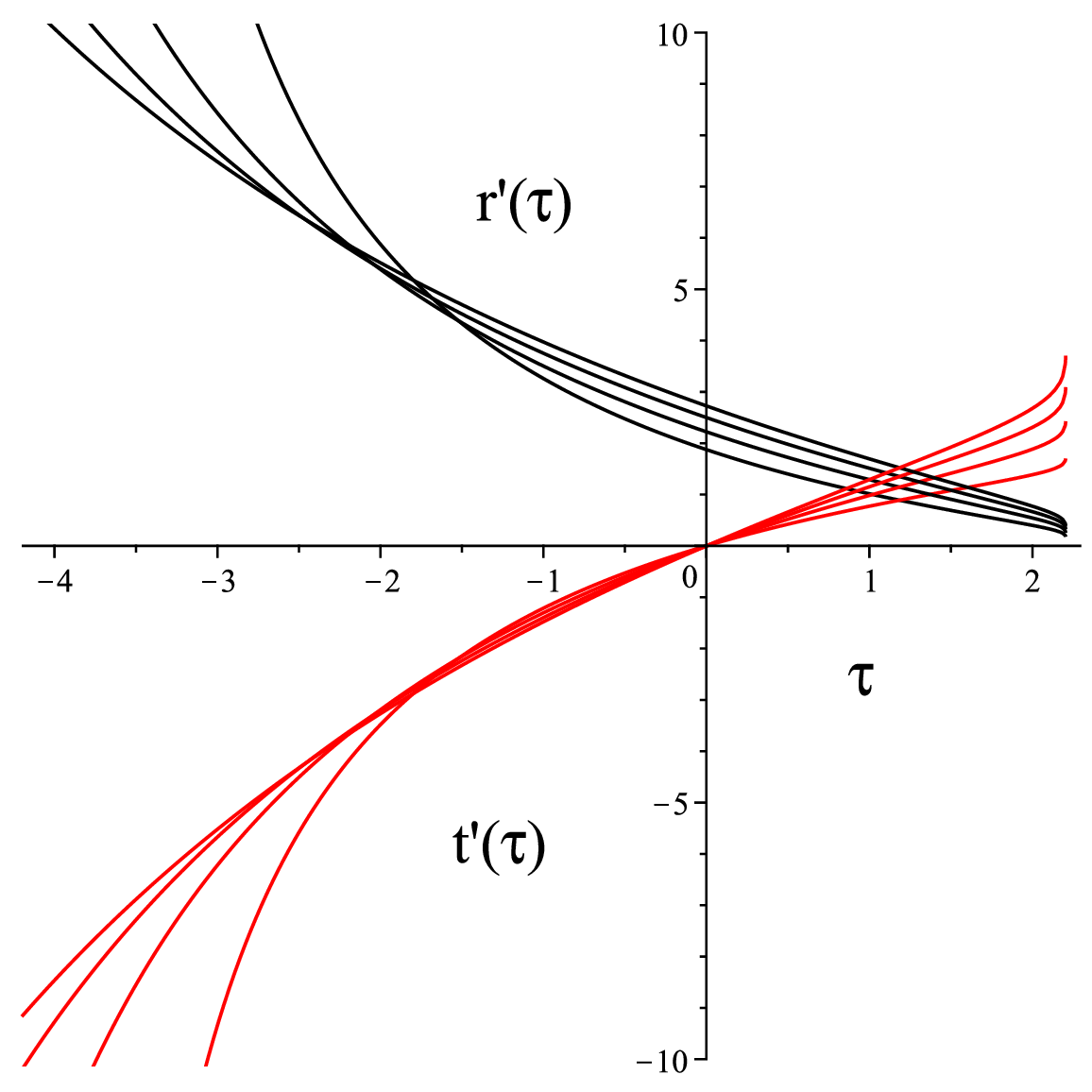}
\caption{\label{fig:2} Upper panel: The orbit of Fig.\ref{fig:1} is 
referred to the projective coordinates $r'$ and $t'$ obtained by 
using the values $A^1{}_1=A^2{}_2=A^0{}_1=A^0{}_2=1$
for the projection matrix, implying
$r' = r/(1 +  t + r)$ (black online) and  
$t' = t/(1 +  t + r)$ (red online). 
However, upon continuing the plots up to $\tau\approx -8.30$ the 
functions $r'(\tau)$ and $t'(\tau)$ diverge, i.e., the map has 
a pole since  we are moving along the geodesics.
Similarly, one cannot extend the plots beyond the value $\tau=2.21$, at which
both $t',r'$ reach their asymptotic values, in this case 
(and within the numerical precision of the plots)
$r'_{\rm asymt} = 0.1079$, $t'_{\rm asymt}= 0.8384$ .
Lower panel: Sequence of plots of $r' = r/(1 +  A^0{}_1t + A^0{}_2r)$ (black online) and  
$t' = t/(1 +  A^0{}_1t + A^0{}_2r)$ (red online) as in the upper panel but 
using different values for $[A^0{}_1,A^0{}_2]=\{[\frac12,\frac13],
[\frac13,\frac14],[\frac14,\frac15],[\frac15,\frac16]\}$.
On the right of $\tau=2$, the numerical integration finds a vertical
asymptote of $t'(\tau)$.}
\end{figure}

\section{Points at infinity in 2-dimensional Schwarzschild spacetime}

Let us first consider a 2-dimensional Schwarzschild spacetime with 
metric written in standard $x^a=(t,r)$ coordinates
(we use $G=c=1$ units)
\begin{equation}
ds^2=-\left(1-\frac{2M}{r}\right)dt^2+\frac{dr^2}{\left(1-\frac{2M}{r}\right)}.
\label{(3.1)}
\end{equation}
It is well known that
\begin{equation}
R_{ab}=\frac{2M}{r^3}g_{ab}\,,\qquad R=\frac{4M}{r^3},
\label{(3.2)}
\end{equation}
i.e.,
\begin{equation}
G_{ab}=R_{ab}-\frac12 R g_{ab}=0\,.
\label{(3.3)}
\end{equation}
The geodesic equation reduces to
\begin{eqnarray}
\label{geo_eqs_sch2d}
\frac{dt}{d\tau} &=& \frac{E }{1-\frac{2M}{r} } , 
\nonumber\\
\frac{dr}{d\tau} &=& \pm\sqrt{ E^2-\epsilon\left(1-\frac{2M}{r}\right)},
\label{(3.4)}
\end{eqnarray}
where $E$ is the energy (Killing) constant and $\epsilon=[1,0,-1]$ for timelike, 
null and spacelike orbits, respectively.

Notice that the $r \to \infty$ limit of the above equation is
\begin{eqnarray}
\frac{dt}{d\tau}=E, 
\nonumber \\
\frac{dr}{d\tau}=\pm \sqrt{E^2 - \epsilon},
\end{eqnarray}
which can be integrated to find
\begin{equation}
t=E \tau , \; 
r=\pm \sqrt{\tau}.
\label{asform}
\end{equation}
In agreement with Sec. II,
we study the following fractional-linear coordinate transformation
\begin{eqnarray}
t'&=&\frac{A^1{}_B y^B}{A^0{}_C y^C}
=\frac{A^{1}{}_{1}t}{\rho},
\nonumber\\
r'&=&\frac{A^2{}_B y^B}{A^0{}_C y^C}
=\frac{A^{2}{}_{2}r}{\rho},
\label{(3.7)}
\end{eqnarray}
where we have defined
\begin{equation}
\rho=A^{0}{}_{0}+A^{0}{}_{1}t+A^{0}{}_{2}r=A^{0}{}_{0}+A^{0}{}_{a}x^a .
\label{(3.8)}
\end{equation}
We can express the differentials as follows:
\begin{eqnarray}
dt' &=& \left(\frac{A^{1}{}_{1}} {\rho}-A^{0}{}_{1}\frac{t'}{\rho}\right)dt
-A^{0}{}_{2}\frac{t'} {\rho}dr ,
\nonumber \\
dr' &=& 
-A^{0}{}_{1}\frac{r'} {\rho}dt
+\left(\frac{A_{\; 2}^{2}}{\rho}
-A^{0}{}_{2}\frac{r'} {\rho}\right)dr ,\qquad
\label{(3.9)}
\end{eqnarray}
implying (concisely, since there is no need to write lengthy expressions explicitly)
\begin{eqnarray}
dt&=& B^t{}_{t'}dt'+B^t{}_{r'}dr',
\nonumber\\
dr&=& B^r{}_{t'}dt'+B^r{}_{r'}dr',
\label{(3.10)}
\end{eqnarray}
and then leading to the new form of the metric
\begin{eqnarray}
ds^2
&=& g_{t't'}dt'{}^2+2g_{t'r'}dt' dr' +g_{r'r'}dr'{}^2.
\label{(3.11)}
\end{eqnarray}
In order to bring infinity down to a finite distance, the choice
of $A$ for which 
\begin{equation}
A^{0}{}_{1} \not = 0, \qquad A^{1}{}_{1} \not = 0,
\label{(3.12)}
\end{equation}
\begin{equation}
A^{0}{}_{2} \not = 0, \qquad A^{2}{}_{2} \not = 0,
\label{(3.13)}
\end{equation}
is of particular interest, because such equations ensure that 
$t\to \infty$ corresponds to 
\begin{equation}
{A^{1}{}_{1}\over A^{0}{}_{1}}
=\lim_{t \to \ \infty} t'
\label{(3.14)}
\end{equation}
at finite values of $r$, and $r=+ \infty$ corresponds to 
\begin{equation}
{A^{2}{}_{2}\over A^{0}{}_{2}}
=\lim_{r \to \infty} r'
\label{(3.15)}
\end{equation}
at finite values of $t$, respectively.

As a first example, we will evaluate the map \eqref{(3.7)} along
the geodesics and exploit the asymptotic limit previously discussed.
Thus, we find the equations 
\begin{eqnarray}
\label{limits_tau_infinity}
\lim_{\tau \to \infty}t'
=\frac{A_{\; 1}^{1}E}
{(A_{\; 1}^{0}E \pm A_{\; 2}^{0}\sqrt{E^2 -\epsilon})},
\nonumber \\
\lim_{\tau \to \infty}r'
=\frac{\pm A_{\; 2}^{2}\sqrt{E^2 -\epsilon}}
{(A_{\; 1}^{0}E \pm A_{\; 2}^{0}\sqrt{E^2 -\epsilon})}.
\end{eqnarray}
At this stage, such limiting values of 
$t'$ and $r'$ can be studied for various choices
of the coefficients of the $A$ matrix, while ruling out the
zeros of the denominator.
Interestingly, the null case ($\epsilon=0$, with $E>0$ for example)  
simplifies further, and leads to
\begin{eqnarray}
\lim_{\tau \to \infty}t'
=\frac{A_{\; 1}^{1}}
{(A_{\; 1}^{0}  \pm A_{\; 2}^{0})},
\nonumber \\
\lim_{\tau \to \infty}r'
=\frac{\pm A_{\; 2}^{2}}
{(A_{\; 1}^{0}  \pm A_{\; 2}^{0})}.
\end{eqnarray}

Let us now consider the numerical integration 
of the timelike geodesics in a 2-dimensional Schwarzschild spacetime, 
examining the radial infalling orbits.
Besides the standard $(t,r)$ coordinates of the metric \eqref{(3.1)}, 
one can describe the radial motion through the projective 
coordinates defined above.
As in Sec. II,  we use the simple choice of $A^B{}_B=1$, $A^0{}_1$ 
and $A^0{}_2$ as only nonvanishing coefficients, implying
\begin{equation}
t'=\frac{t}{\rho},\qquad r'=\frac{r}{\rho},
\label{(3.18)}
\end{equation}
with $\rho=1+tA^0{}_1+rA^0{}_2$.
We have then (explicitly, in this special case)
\begin{eqnarray}
dt'&=&\frac{1}{\rho}\left(1-\frac{tA^0{}_1}{\rho}\right)dt-\frac{t}{\rho^2}A^0{}_2dr,
\nonumber\\
dr'&=& -\frac{r}{\rho^2}A^0{}_1dt+\frac{1}{\rho}\left(1-\frac{rA^0{}_2}{\rho}\right)dr,
\label{(3.19)}
\end{eqnarray}
with inverse relations (identifying the coefficients $B^i{}_j$ of Eq. \eqref{(3.10)})
\begin{eqnarray}
dt&=& \rho^2 (- r' A^0{}_2 +1) dt'+ \rho^2 t' A^0{}_2 dr' , 
\nonumber\\
dr&=&  \rho^2  r'A^0{}_1 dt' + \rho^2 (-t'A^0{}_1 +1) dr' ,
\label{(3.20)}
\end{eqnarray}
and the new metric \eqref{(3.11)} having components
\begin{eqnarray}
g_{t't'}&=&-\frac{\rho^3(\rho r'- 2 M)(r' A^0{}_2  - 1)^2}{r'}   
+ \frac{\rho^5 r'{}^3(A^0{}_1)^2 }{ (\rho r' - 2M)}, 
\nonumber\\
g_{t'r'}&=&   \frac{(\rho r' - 2M)\rho^3 (r' A^0{}_2  - 1) t' A^0{}_2}{r'}
\nonumber\\ 
&&-\frac{  r'{}^2A^0{}_1 (t' A^0{}_1 - 1)\rho^5}{(\rho r' - 2M)}, 
\nonumber\\
g_{r'r'}&=& -\frac{(\rho r' - 2 M)\rho^3 t'{}^2(A^0{}_2)^2}{r'}
\nonumber\\ 
&&+ \frac{\rho^5 (t' A^0{}_1 - 1)^2 r'}{ (\rho r' - 2M)}.
\label{(3.21)}
\end{eqnarray}

With a simple choice of parameters and initial conditions the 
numerical integration of the geodesics 
with $(t,r)$ coordinates is summarized in Fig. 
\ref{fig:1}. Figure \ref{fig:2} offers instead the picture of 
the same geodesic orbit but with the projective coordinates $(t',r')$.
We have chosen therein rational values of $A_{\; 1}^{0}$ and
$A_{\; 2}^{0}$ so that, by virtue of \eqref{(3.14)} and \eqref{(3.15)},
timelike and spacelike infinity read as
$$
(2,0),(3,0),..., (0,3),(0,4),...
$$
respectively. 

For this simple example we can consider the observers adapted to the 
projective coordinates, Eq. \eqref{(3.18)}.
They form a congruence of accelerated worldlines, with nonzero expansion but 
vanishing vorticity. The use of such observers as fiducial observers will anyway 
complicate matters. For example, together with
\begin{equation}
u_{\rm pc}=\frac{1}{\sqrt{-g_{t't'}}}
\frac{\partial}{\partial t'}
\label{(3.22)}
\end{equation}
one has the orthogonal direction
\begin{equation}
e(u_{\rm pc})_1=Z_{1} \left[g_{r't'}\frac{\partial}{\partial t'}
-g_{t't'}\frac{\partial}{\partial r'}\right]
\label{(3.23)}
\end{equation}
with $Z_{1}$ obtained from the normalization condition, 
\begin{equation}
Z_{1} =\frac{1}{\sqrt{-g_{t't'}}\sqrt{ (g_{r't'})^2-g_{t't'}  g_{r'r'} }}.
\label{(3.24)}
\end{equation}
The acceleration of $u_{\rm pc}$ is then
\begin{equation}
a(u_{\rm pc})=\nabla_{u_{\rm pc}} u_{\rm pc}=\kappa e(u_{\rm pc})_1 .
\label{(3.25)}
\end{equation}
Here $\kappa$ is a complicated function of $t'$ and $r'$.
For example, for $A^0{}_1=1$, $A^0{}_2=0$ we find
\begin{equation}
\kappa^2=M^2 \frac{(1-t')^4 Z_3(Z_{2})^{2}
}{ r'{}^3 (Z_3^2-r'{}^4)^3},
\label{(3.26)}
\end{equation}
where  
\begin{equation}
Z_3=r'+2M (t'-1), \;
Z_{2}=-3 r'{}^4+Z_1^2 .
\label{(3.27)}
\end{equation}

We can also consider in this specific case the slicing observers, defined in Eq. \eqref{eq_sli_obs}.
They have their four-velocity vector orthogonal to the $t'=$ constant hypersurfaces 
and form a congruence of (radially) accelerated worldlines. In the specific case discussed above we find
\begin{equation}
-g^{t't'}=\frac{f_1 f_2}{\rho^5 (-r'\rho+2 M)r'},
\end{equation}
where
\begin{eqnarray}
f_1&=& (-r'A^0{}_2 (-t'+r') \rho^2+r'(-1+2 M A^0{}_2) \rho+2 M)
\nonumber\\
f_2&=&  (-r'A^0{}_2 (t'+r')\rho^2+r'(-1+2 M A^0{}_2) \rho
\nonumber \\
&+& 2 M).
\end{eqnarray}

\begin{figure}
\includegraphics[scale=0.3]{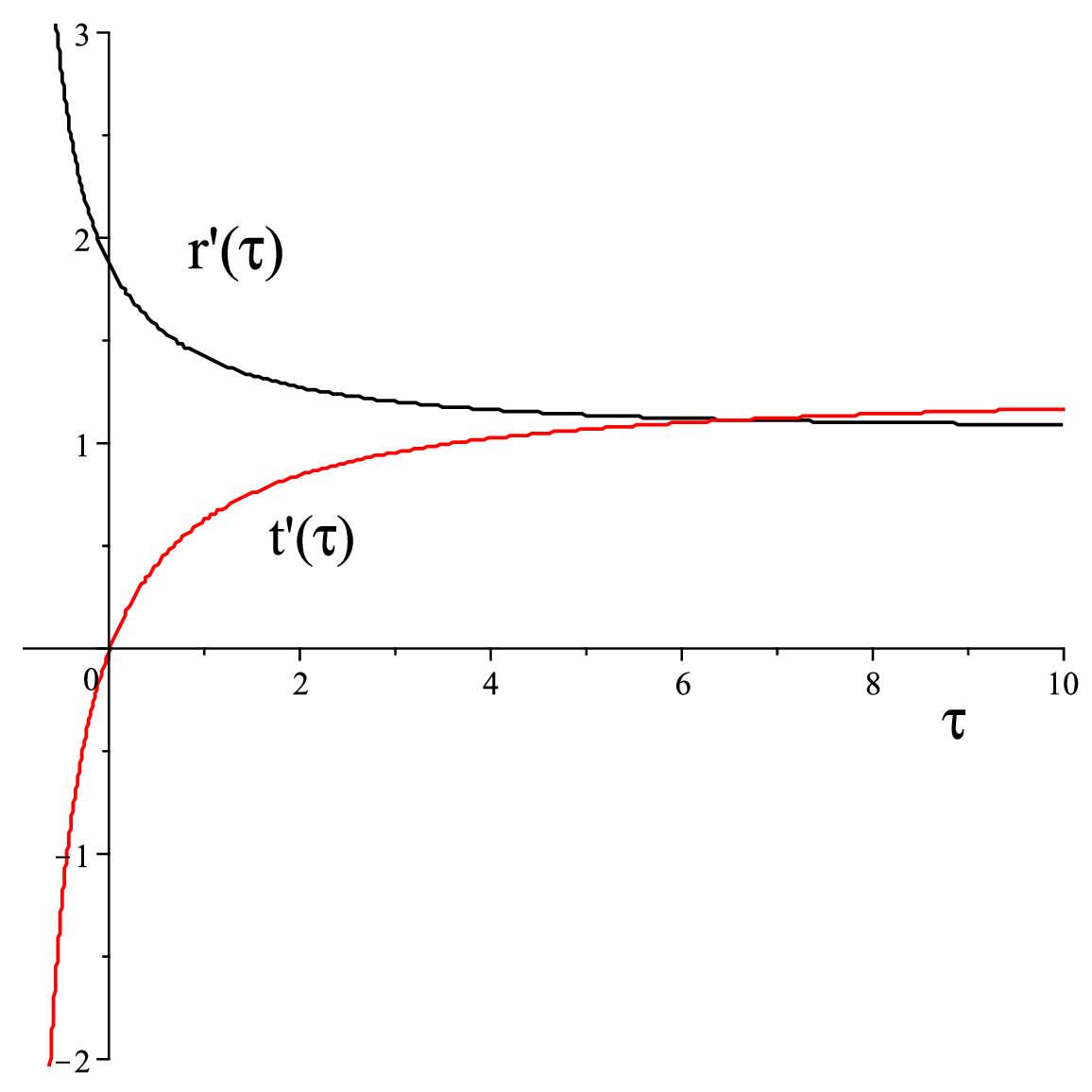}\\
\includegraphics[scale=0.3]{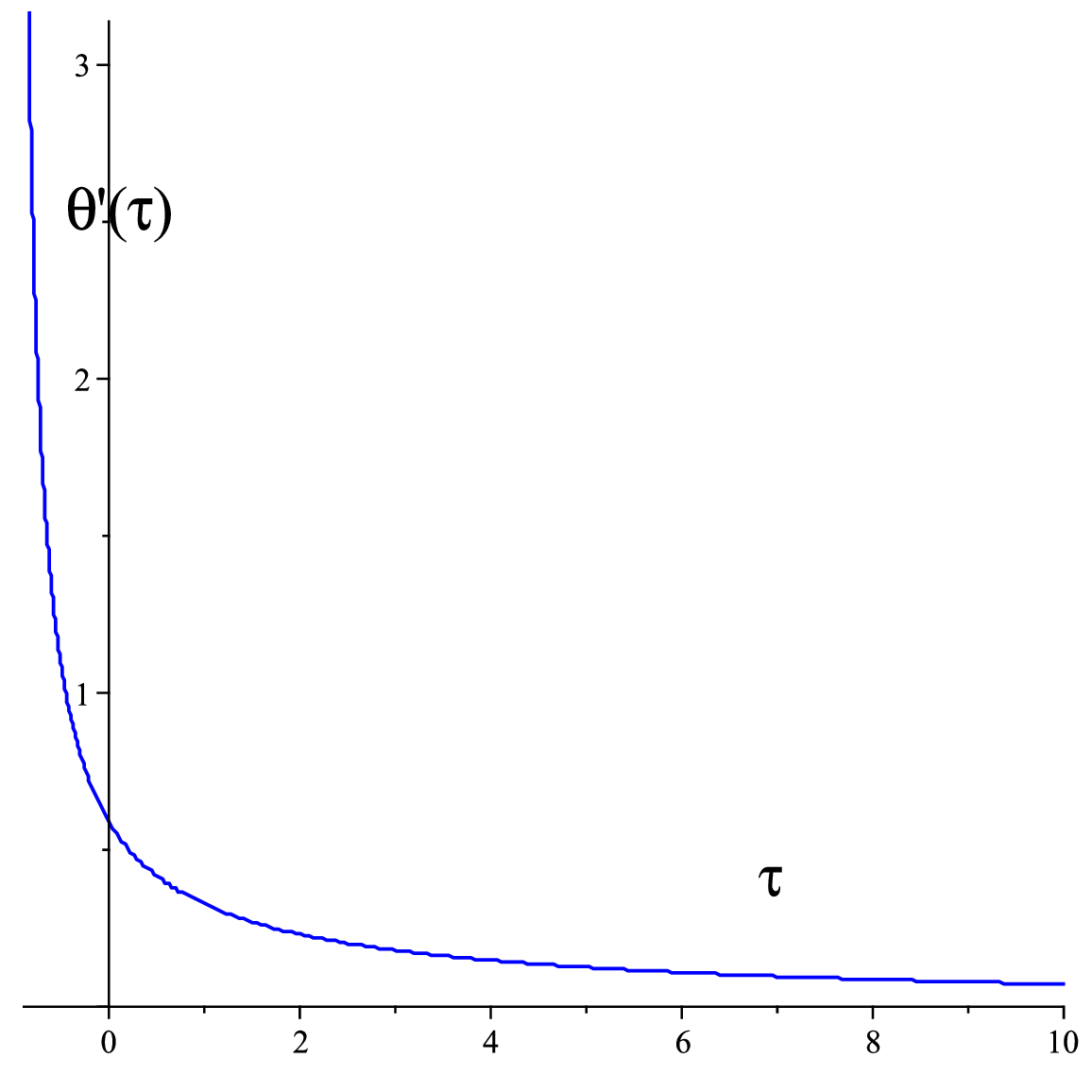}\\
\includegraphics[scale=0.3]{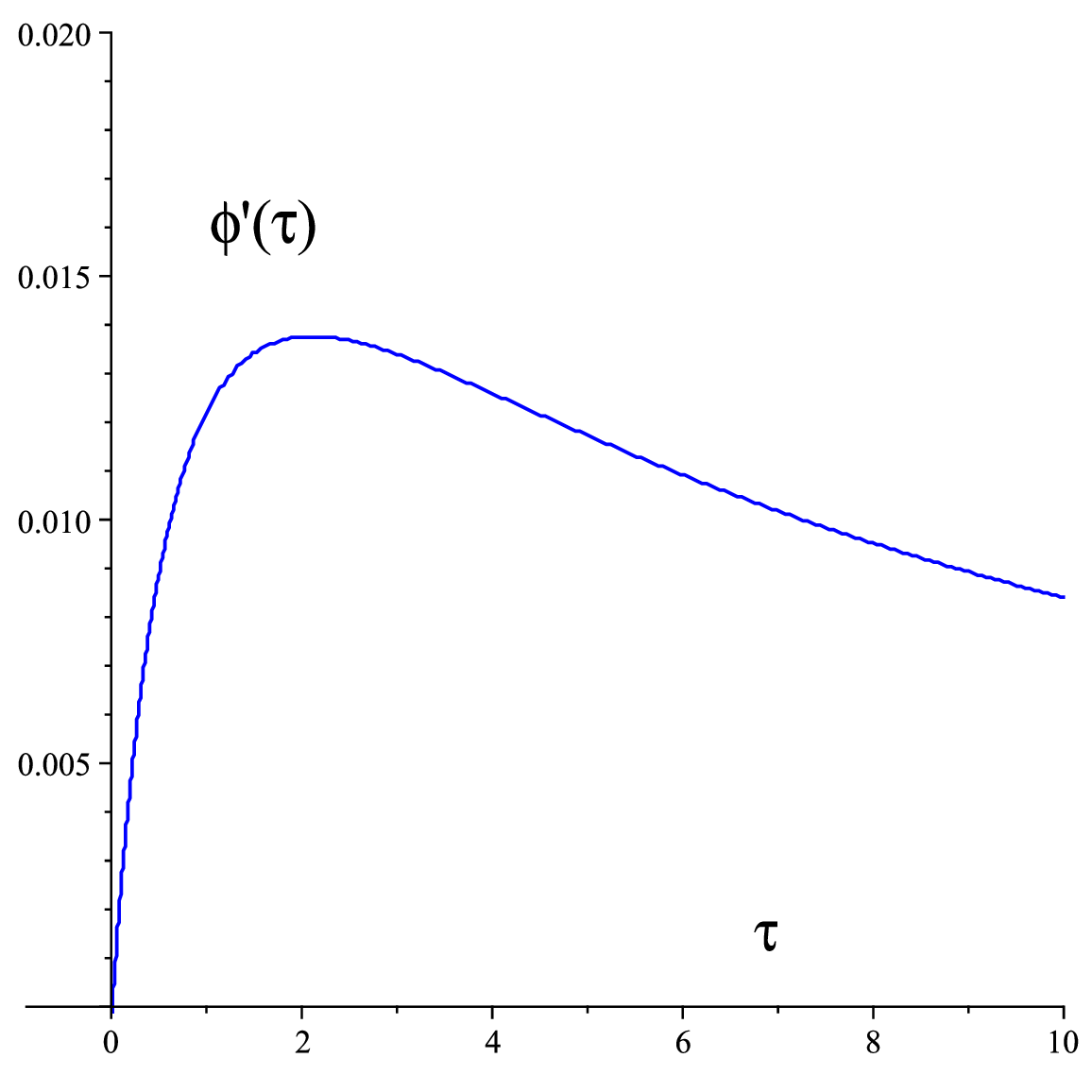}\\
\caption{\label{fig:x} Timelike equatorial plane geodesics of a 
four-dimensional Scwarzschild spacetime mapped into projective coordinates.
$t' =t/(1 + A^0{}_1t  + A^0{}_2 r)$ (red online), 
$r' = r/(1 + A^0{}_1t  + A^0{}_2 r)$ (black online) 
$\theta' = \pi/2(1 + A^0{}_1t  + A^0{}_2 r)$ (grey online) 
$\phi' = \phi/(1 + A^0{}_1t  + A^0{}_2 r)$ (blue online). 
The original Schwarzschild geodesics are radially outgoing with parameters
$M=1$,$E=2$ (energy per unit of mass), $L=2$  (angular momentum per unit  of mass) 
following the initial conditions $\phi(0) = 0, r(0) = 5, t(0) = 0$.
The panels a), b), c) refer to $r'(\tau)$, $\theta'(\tau)$, $\phi'(\tau)$ 
respectively. Note the plot ranges of the transformed angles $\theta'\in[0,\pi[$ and $\phi'\in[0,2\pi[$.
The projective parameters chosen are $[A^0{}_1,A^0{}_2]=[\frac12,\frac13]$.
This projective picture is relevant, for example, for 
well known applications of the Schwarzschild spacetime, including for example  all timelike geodesics 
which describe the motion of planets about the Sun.  
The same qualitative behavior is obtained on considering other
choices of $A_{\; 1}^{0}$ and $A_{\; 2}^{0}$, as well as on studying
the projective version of null geodesics. Thus, we have avoided
displaying multiple figures with the same content.} 
\end{figure}

\begin{figure}
\includegraphics[scale=0.3]{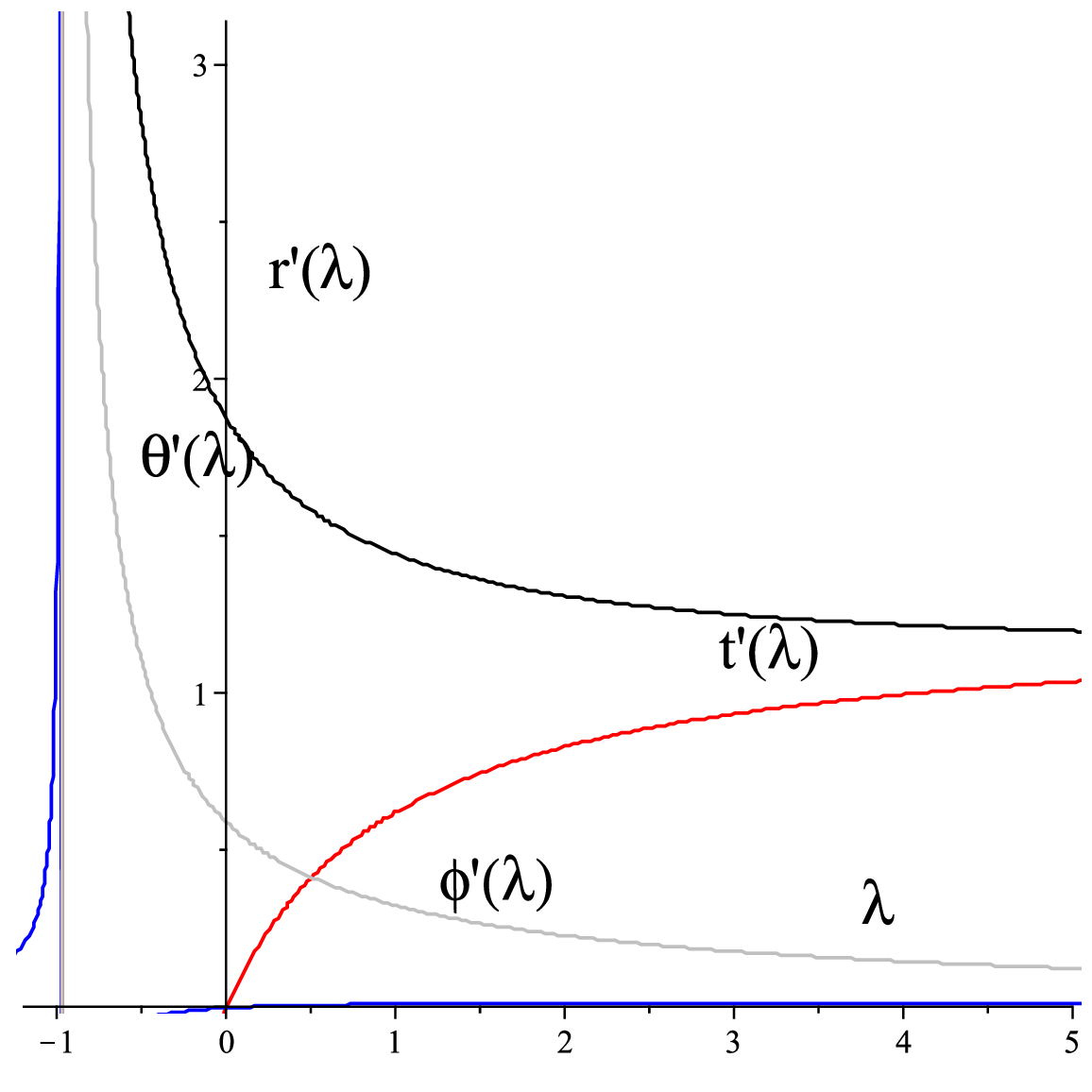}
\caption{\label{fig:xbis} By using the same parameters and initial conditions of 
Fig. \ref{fig:x} we display the behavior of the null equatorial plane 
geodesics as functions of an affine parameter $\lambda$.}
\end{figure}

\section{Second look at points at infinity in Schwarzschild spacetime}

In order to further test the method of Sec. II, 
we here consider the standard $(t,r,\theta,\phi)$
coordinates for Schwarzschild geometry, for which, upon exploiting
a ${\rm GL}(5,{\mathbb R})$ matrix with components
$A_{\; C}^{B}$, we pass to new coordinates 
\begin{eqnarray}
t'&=& \frac{T}{\rho},\qquad
r'= \frac{R}{\rho},
\nonumber\\
\theta'&=& \frac{\Theta}{\rho}\,,\qquad
\phi'= \frac{\Phi}{\rho},
\label{(4.1)}
\end{eqnarray}
where
\begin{eqnarray}
T(t,r,\theta,\phi)
&=& A^{1}{}_{0}+A^{1}{}_{\alpha }x^\alpha
\nonumber\\
R(t,r,\theta,\phi)
&=& A^{2}{}_{0}+A^{2}{}_{\alpha }x^\alpha
\nonumber\\
\Theta(t,r,\theta,\phi)
&=& A^{3}{}_{0}+A^{3}{}_{\alpha }x^\alpha
\nonumber\\
\Phi(t,r,\theta,\phi)
&=& A^{4}{}_{0}+A^{4}{}_{\alpha }x^\alpha ,
\label{(4.2)}
\end{eqnarray}
with 
\begin{eqnarray}
\rho(t,r,\theta,\phi) &=& {{y'}^{0}\over y^{0}}
=\sum_{\nu=0}^{4}A^{0}{}_{\nu}{y^{\nu}\over y^{0}}
\nonumber \\
&=& A^{0}{}_{0}+A^{0}{}_{\alpha}x^\alpha.
\label{(4.3)}
\end{eqnarray}
Hereafter, we assume the form \eqref{(2.17)} for the matrix $A$ 
for the reasons described in Sec. II, and hence we find
\begin{eqnarray}
dt'&=&\frac{1}{\rho}\left(1-\frac{A_{\; 1}^{0}t} {\rho}
\right)dt-\frac{A_{\; 2}^{0}t}{\rho^{2}}dr,
\nonumber\\
dr'&=&-\frac{A_{\; 1}^{0}r}{\rho^{2}}dt
+\frac{1}{\rho}\left(1-\frac{A_{\; 2}^{0}r}{\rho}
\right)dr,
\nonumber\\
d\theta'&=&-\frac {A_{\; 1}^{0}\theta}{\rho^{2}}dt
-\frac{A_{\; 2}^{0}\theta}{\rho^{2}}dr
+\frac{1}{\rho}d\theta,
\nonumber\\
d\phi'&=&-\frac {A_{\; 1}^{0}\phi}{\rho^{2}}dt
-\frac{A_{\; 2}^{0}\phi}{\rho^{2}}dr
+\frac{1}{\rho}d\phi.
\label{(4.4)}
\end{eqnarray}
At this stage, we solve the system \eqref{(4.4)} 
for $dt,dr,d\theta$ and $d\phi$ in the form
\begin{equation}
dx^{\mu}=B^{\mu}{}_{\nu}(x')d{x'}^{\nu},
\label{(4.5)}
\end{equation}
bearing in mind that
\begin{equation}
t=\rho t', \; r=\rho r', \; 
\theta=\rho \theta', \; \phi=\rho \phi',
\label{(4.6)}
\end{equation}
while $\rho$ obeys the equation
\begin{equation}
\rho=1+A_{\; 1}^{0}t+A_{\; 2}^{0}r
=1+\rho(A_{\; 1}^{0}t'+A_{\; 2}^{0}r'),
\label{(4.7)}
\end{equation}
which is solved by
\begin{equation}
\rho=\left(1-A_{\; 1}^{0}t'-A_{\; 2}^{0}r'\right)^{-1}
=\rho(t',r').
\label{(4.8)}
\end{equation}
Hence we find that the metric has nondiagonal form in the
$(t',r',\theta',\phi')$ coordinates:
\begin{widetext}
\begin{eqnarray}
g_{t't'} &=& \frac{\rho^3}{r'} F(r'A^0{}_2 - 1)^2   
- \frac{r'{}^3\rho^5 \Bigr(A_{\; 1}^{0}\Bigr)^2}{F} 
+ \Bigr(A_{\; 1}^{0}\Bigr)^2 \rho^6 r'{}^2 \Omega'
\nonumber\\
g_{r'r'} &=& \frac{\rho^3}{r'} F t'{}^2 \Bigr(A_{\; 2}^{0}\Bigr)^2  
- \frac{r'\rho^5 (-t'A^0{}_1 + 1)^2}{F} 
+ \Bigr(A_{\; 2}^{0}\Bigr)^2\rho^6 r'{}^2 \Omega'
\nonumber\\
g_{\theta'\theta'} &=& r'{}^2\rho^4
\nonumber\\
g_{\phi' \phi'} &=& \rho^4{\mathcal R}^2 
\nonumber\\
g_{t' r'} &=& -\frac{\rho^3}{r'}F (r' A^0{}_2 - 1) t' A^0{}_2  
- \frac{r'{}^2\rho^5 A^0{}_1(-t'A^0{}_1 + 1)}{F} 
+ A^0{}_1 A^0{}_2\rho^6 r'{}^2 \Omega'
\nonumber\\
g_{t' \theta'} &=&\rho^5 r'{}^2 \theta' A^0{}_1
\nonumber\\
g_{t'\phi'} &=&  \rho^5{\mathcal R}^2 \phi' A^0{}_1
\nonumber\\
g_{r'\theta'} &=& \rho^5 r'{}^2\theta' A^0{}_2
\nonumber\\
g_{r'\phi'} &=& \rho^5{\mathcal R}^2 \phi' A^0{}_2
\nonumber\\
g_{\theta'\phi'} &=& 0,
\label{(4.9)}
\end{eqnarray}
where
\begin{equation}
\Omega'=\sin^2(\theta'\rho)\phi'{}^2 + \theta'{}^2 , \qquad 
F=2M -r'\rho, \qquad {\mathcal R}^2=r'{}^2 \sin^2(\theta'\rho).
\label{(4.10)}
\end{equation}
\end{widetext}
This is a particular case of the spherically symmetric metric 
with inhomogeneous projective coordinates obtained by us in Appendix C. 

We now describe briefly the projective version of timelike
and null geodesics by using Eq. \eqref{(2.16)}, and then using for
$t,r,\theta,\phi$ on the right-hand side therein the formulas obtained
in Ref. \cite{Chandra}, limiting to equatorial plane 
$\theta=\pi/2$ geodesics without any loss of generality.
For this purpose it is worth recalling shortly here the expressions 
of the parametric equations $x^\mu =x_p^{\mu}(\tau)$ for generic geodesics 
in the original Schwarzschild coordinates, limiting ourselves for 
simplicity to the equatorial plane (Figs. 3 and 4)
\begin{eqnarray}
\label{geoeqnsschw}
\frac{dt_p}{d\tau}&=&\frac{E}{f(r_p)},
\nonumber\\
\frac{dr_p}{d\tau}&=&\epsilon_r\left[E^2-f(r_p)
\left(\epsilon+\frac{L^2}{r_p^2}\right)\right]^{1/2},
\nonumber\\
\frac{d\phi_p}{d\tau}&=&\frac{L}{r_p^2},
\end{eqnarray}
where $f=1-\frac{2M}{r}$, 
$\epsilon_r=\pm1$ is a sign indicator keeping track of 
increasing/decreasing radial coordinate, and $E$ and $L$ denote the 
particle's energy and angular momentum per unit mass, respectively.
At this stage, one can proceed in a way entirely analogous to what we
have done in Eqs. \eqref{limits_tau_infinity}.

\subsection{Scalar field in Schwarzschild: standard vs projective coordinates}

In this brief subsection we aim at outlining the projective description
of an important topic in gravitational radiation and black-hole 
physics, i.e., the scalar wave equation for a massless scalar field
in a fixed Schwarzschild background. For this purpose, we write 
the spherically symmetric wave equation
\begin{equation}
\label{eq_fund}
\Box \psi=0,
\end{equation}
where
\begin{equation}
\Box \psi=\frac{1}{\sqrt{-g}}\partial_\mu (\sqrt{-g}g^{\mu\nu}\partial_\nu  \psi),
\label{(2)}
\end{equation}
on a fixed Schwarzschild background with mass $M$. 
By virtue of the symmetries of 
Schwarzschild spacetime in standard spherical coordinates, 
the solution of the scalar wave equation reads as
\begin{equation}
\label{psi_sol}
\psi(t,r,\theta,\phi)=\sum_{lm}\int \frac{d\omega}{2\pi}e^{-i\omega t}R_{lm\omega}(r)Y_{lm}(\theta,\phi),
\end{equation}
where $Y_{lm}(\theta,\phi)$ are the (scalar) spherical harmonics and the radial 
function $R_{lm\omega}(r)$ satisfies the homogeneous equation
(setting $L=l(l+1)$, $f(r)=1-\frac{2M}{r}$)
\begin{equation}
\label{rad_eq}
R_{lm\omega}''+\frac{2(r-M)}{r^2f(r)}R_{lm\omega}'+\frac{1}{f}\left(\frac{\omega^2}{f}  
- \frac{L}{r^2}\right)R_{lm\omega}=0,
\end{equation}
which is solved in terms of confluent Heun functions.
It is well known that separation of variables is a matter of coordinates. In fact 
the latter property is lost when passing to the projective coordinates.
For example, even the special case \eqref{(4.6)}, where $\rho$ is given in
Eq. \eqref{(4.8)}, implies that 
\begin{eqnarray}
\label{psi_sol}
\tilde \psi (t',r',\theta',\phi')&=&\psi(t,r,\theta,\phi)\big|_{x^\alpha =\rho x'^\alpha}\nonumber\\
&=& \sum_{lm}\int \frac{d\omega}{2\pi}e^{-i\omega \rho t'}R_{lm\omega}(\rho r')\times \nonumber\\
&& \times Y_{lm}(\rho \theta',\rho \phi'),
\end{eqnarray}
i.e., no simple relations between $\psi(t,r,\theta,\phi)$ and its transformed 
version exists (and no separation of variables is available).
The advantage of using projective-type coordinates, as expected, arises when looking 
at asymptotic behaviors in $(t,r,\theta,\phi)$  which correspond to behaviors 
at finite values of $(t',r',\theta',\phi')$.
However, even in this case no special simplifications seem to occur.

In fact, let us consider the $l=0=m$ mode so that the dependence on the angular 
variables is frozen, and the soft-frequency limit $\omega \to 0$ can be taken.
In this case, in the original Schwarzschild variables 
the radial function $R(r)=R_{0,0,0}(r)$ satisfies the equation
\begin{equation}
\label{rad_eq_simp}
R''+\frac{2(r-M)}{r^2f(r)}R'=0,
\end{equation}
with solution
\begin{equation}
R(r)=\frac{c_1}{2M}\ln f(r) +c_2 .
\end{equation}
We can set $c_2=0$ to ensure $R(r)\to 0$ when $r\to \infty$. We also choose $c_1$ so that
\begin{equation}
\frac{1}{\sqrt{4\pi}} \frac{c_1}{2M}=1.
\end{equation}
We obtain therefore
\begin{equation}
\tilde \psi (t',r',\theta',\phi')\big|_{l=0=m}=\ln\left(1-\xi(t',r')  \right),
\end{equation}
with
\begin{equation}
\xi(t',r')=\frac{2M(1-A^0{}_1t'- A^0{}_2 r')}{r'} . 
\end{equation}
The limit $r'\to \infty$  (at fixed value of $t'$) leads to the constant value
\begin{equation}
\tilde \psi (t',r',\theta',\phi')\big|_{l=0=m}=\ln\left(1+ 2M A^0{}_2 \right), 
\end{equation}
where we recall that the projective parameters have originally the dimension of a length, 
whereas for simplicity we resort to dimensionless coordinates 
in the whole paper. Measuring them in terms of the 
natural length scale of this case, $M$, we would have $\tilde A^0{}_2=MA^0{}_2$.

Interestingly, the asymptotic form of $\psi$ is fully computable despite the 
technical complications resulting from a non-diagonal form of the metric.
Thus, one can envisage a long-term research program devoted to wave equations
in projective coordinates for spherically symmetric spacetimes.

\section{Nariai spacetime}

We now consider the Nariai spacetime model \cite{Nariai1,Nariai2}
which is of particular interest
because for it the Penrose conformal boundary cannot be defined.
Following Ref. \cite{Kroon} one can point out that, if a solution
of the Einstein equations admits a smooth conformal extension,
then the conformally rescaled Weyl tensor should satisfy
the condition
$$
{\widetilde C}_{\lambda \mu \nu \rho}
{\widetilde C}^{\lambda \mu \nu \rho}=0.
$$
But in the Nariai case, one finds
$$
{\widetilde C}_{\lambda \mu \nu \rho}
{\widetilde C}^{\lambda \mu \nu \rho}
={\rm constant} \;  \not =0.
$$
This contradicts the above condition, and hence there
cannot exist a piece of conformal boundary which is
of class $C^{2}$ \cite{Kroon}.

The Nariai spacetime metric \cite{Nariai1,Nariai2} written in dimensionless 
spherical-like coordinates $(t,r,\theta,\phi)$ reads 
as\footnote{Dimensionful coordinates $[T,R]$ can be restored by 
using $\Lambda\sim 1/L^2$ as an overall length scale, that is
$$
ds^2=-dT^2+ \cosh^2 (T\sqrt{\Lambda})\, dR^2 
+\frac{1}{\Lambda}[d\theta^2+\sin^2\theta d\phi^2].
$$
This metric solves Einstein's equation $G_{\mu\nu}+\Lambda g_{\mu\nu}=0$, 
i.e., for any value of $\Lambda$.}
\begin{equation}
ds^2=-dt^2+\cosh^2 t\, dr^2 +d\theta^2+\sin^2\theta d\phi^2,
\label{(5.1)}
\end{equation}
and is an exact solution of Einstein's field equation with 
cosmological constant $\Lambda=-1$, i.e., with
\begin{equation}
G_{\alpha \beta}=g_{\alpha\beta}.
\label{(5.2)}
\end{equation}
It has several important properties \cite{Kroon}:
\begin{enumerate}
  \item geodesic completeness \cite{Beyer};
  \item globally hyperbolicity;
  \item it does not even admit a patch of a conformal boundary;
  \item  it is algebraically special of Petrov type D.
\end{enumerate}
The $t-r$ part of the metric 
\begin{equation}
ds_{(t,r)}^2=-dt^2+\cosh^2 t\, dr^2\,,
\end{equation}
can be used to visualize the modifications to the light cones as 
time evolves. For instance, looking at the family of points
$(r_0,t_0)$ with $r_0$ fixed and $t_0$ taken as a parameter, 
the light-cone equation becomes
\begin{equation}
t-t_0=\pm (\cosh  t_0) (r-r_0).
\end{equation}
This relation leads to the $45$ degree opening angle at $t_0=0$, and 
then the angle gets restricted continuously, vanishing in the 
limit $t_0\to \infty$. Differently, with $t_0$ fixed and $r_0$ taken 
as a parameter, the light-cone structure is fixed itself, i.e., the 
opening angle does not depend on $r_0$.

The spherical-like coordinates are adapted to the two Killing 
vector fields
\begin{equation}
\frac{\partial}{\partial r},
\qquad 
\frac{\partial}{\partial \phi}.
\label{(5.3)}
\end{equation} 
The existence of these Killing symmetries makes it possible to separate the 
geodesics. On denoting their parametric equations by $x^\alpha=x^\alpha(\tau)$,
one has (Figs. 5 and 6)
\begin{eqnarray}
\frac{dt}{d\tau}&=&  \sqrt{C_2+\frac{C_1^2}{\cosh^2 t }},
\nonumber\\
\frac{dr}{d\tau}&=& \frac{C_1}{\cosh^2 t }\,,\nonumber\\
\frac{d\theta}{d\tau}&=& \pm \sqrt{C_3-\frac{L^2}{\sin^2\theta}},
\nonumber\\
\frac{d\phi}{d\tau}&=&\frac{L}{\sin^2 \theta } ,
\label{(5.4)}
\end{eqnarray}
where the $C_i$ and $L$ are constant and we have chosen 
$dt/d\tau>0$ to have future-oriented orbits.
Interestingly, these equations are pairwise coupled, i.e., the coupled 
variables are $(t,r)$ (corresponding to geodesic motion on the 
pseudosphere) and $(\theta, \phi)$ (corresponding to geodesic 
motion on the $2$-sphere), with no mutual intersections.
For instance, by assuming $L=0$ and $C_3=0$ one has $\theta(\tau)
=\theta(0)$, $\phi(\tau)=\phi(0)$ and one is left with the temporal 
and radial equations only.  
The normalization condition for these orbits: $U\cdot U=-\epsilon$, 
with $U^\alpha=\frac{dx^\alpha}{d\tau}$ and $\epsilon=[1,0,-1]$ for timelike, 
null, spacelike orbits, respectively, reads as
\begin{equation}
\label{normaliz}
-\left(\frac{dt}{d\tau}\right)^2+\left(\frac{d\theta}{d\tau}\right)^2
+\frac{C_1^2}{\cosh^2 t }+\frac{L^2}{\sin^2 \theta }+\epsilon=0,
\end{equation}
where only the Killing relations have been used. Inserting in Eq. 
\eqref{normaliz} the full set of Eqs. \eqref{(5.4)} one finds
\begin{equation}
C_2=C_3+\epsilon,
\end{equation}
meaning that, once a causality condition is imposed, the integration 
constants $C_2$ and $C_3$ are no longer freely specifiable.

\begin{figure}
\includegraphics[scale=0.3]{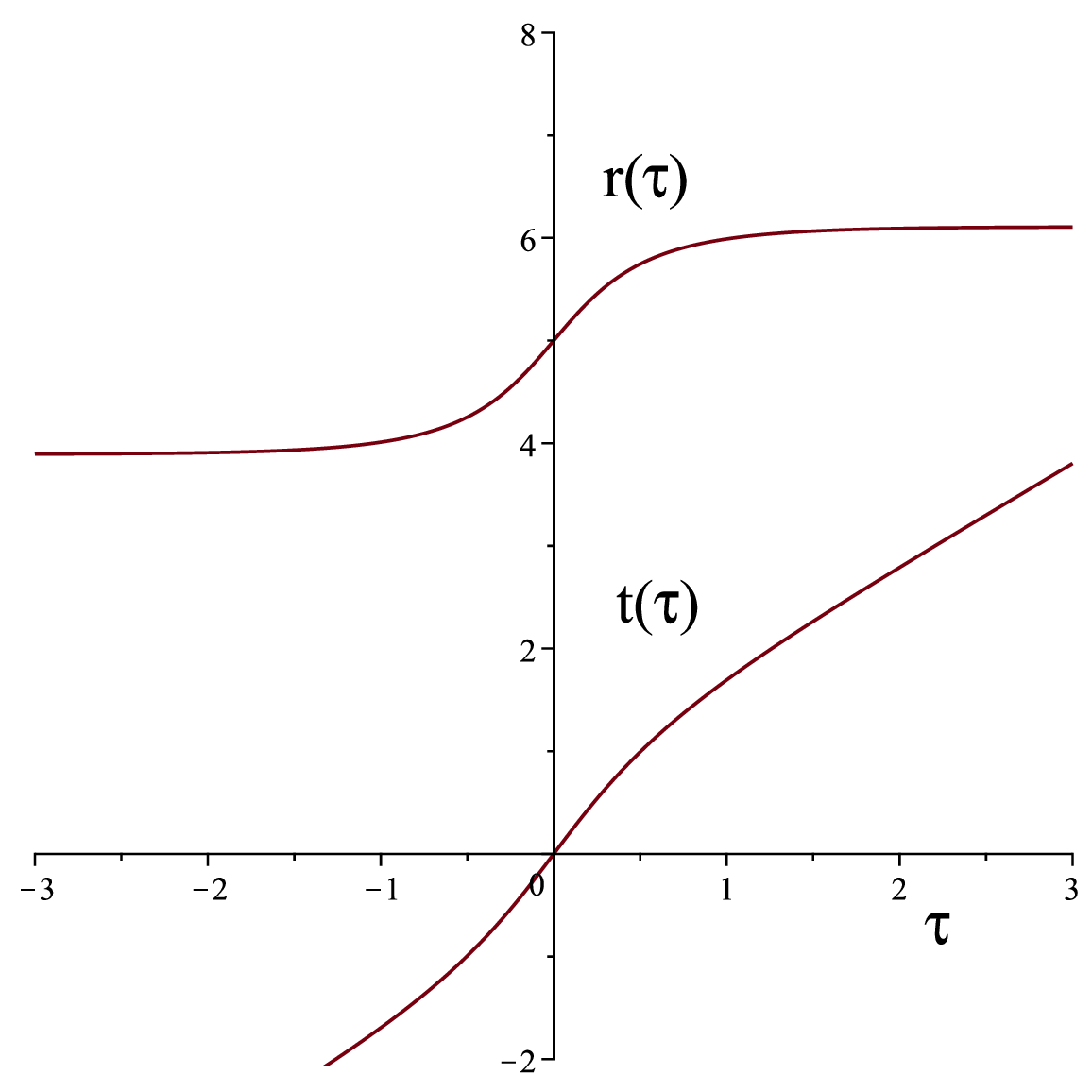}
\includegraphics[scale=0.3]{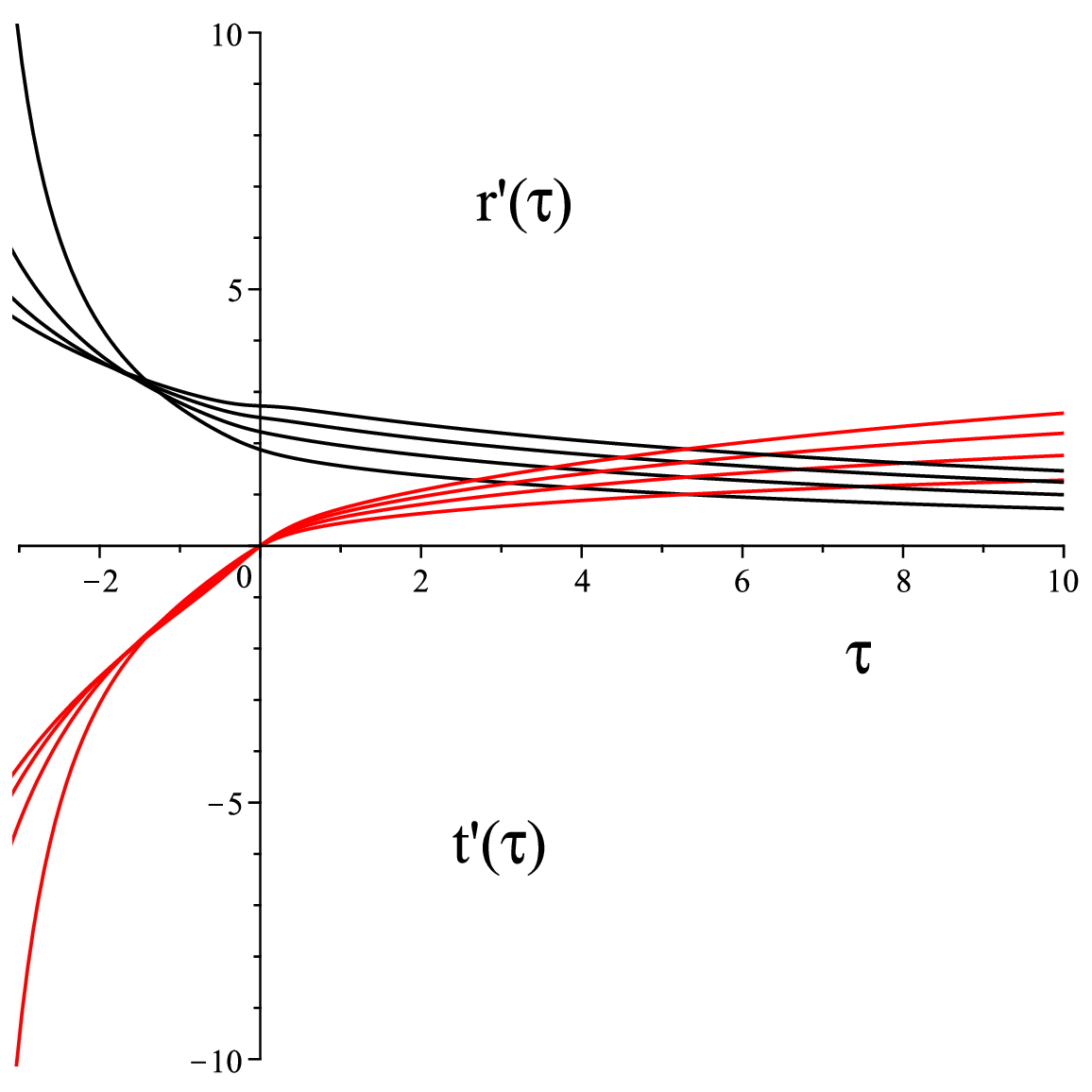}
\caption{\label{fig:5}  Upper panel:  Example of numerical integration 
of the timelike geodesics in the Nariai spacetime with parameters
 $C_1 = 2, C_2 = 1, C_3 = 0, L = 0$
and initial conditions $\phi(0) = 0, r(0) = 5, t(0) = 0, \theta(0) = \frac{\pi}{4}$.
Lower panel: The same map as in Fig. \ref{fig:2} (lower panel), i.e.,
$r' = r/(1 + A^0{}_1t  + A^0{}_2 r)$ (black online) and  
$t' =t/(1 + A^0{}_1t  + A^0{}_2 r)$ (red online), 
using different values for $[A^0{}_1,A^0{}_2]=\{[\frac12,\frac13],
[\frac13,\frac14],[\frac14,\frac15],[\frac15,\frac16]\}$.} 
\end{figure}

\begin{figure}
\includegraphics[scale=0.3]{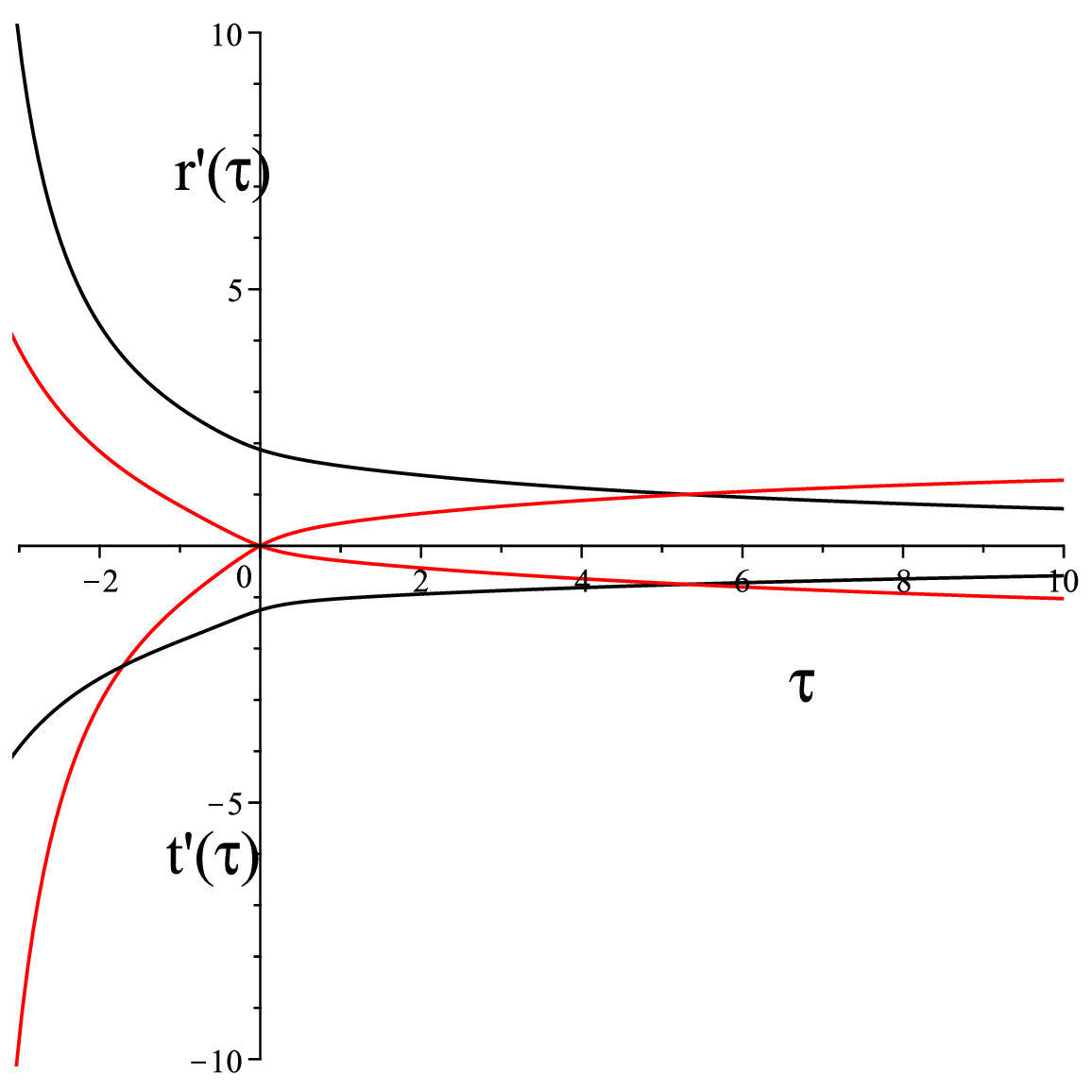}
\caption{The lower panel of Fig. 5 repeated by using  
for $A^0{}_1=\frac{1}{n},A^0{}_2=\frac{1}{(n+1)}$ for $n=2,-2$.}
\end{figure}

A natural orthonormal frame is the following:
\begin{eqnarray}
e_0&=& \frac{\partial}{\partial t},
\qquad e_1
=\frac{1}{\cosh t} \frac{\partial}{\partial r},
\nonumber\\
e_2&=& \frac{\partial}{\partial \theta},
\qquad e_3=\frac{1}{\sin\theta} \frac{\partial}{\partial \phi},
\label{(5.5)}
\end{eqnarray}
which can be used to form a standard Newman-Penrose frame
\begin{eqnarray}
l&=&\frac{1}{\sqrt{2}}(e_0+e_2)\,,\qquad n=\frac{1}{\sqrt{2}}(e_0-e_2),
\nonumber\\
m&=&\frac{1}{\sqrt{2}}(e_1+ie_3).
\label{(5.6)}
\end{eqnarray}
The non-vanishing spin coefficients reduce to
\begin{eqnarray}
\sigma&=&-\mu
=\frac{\sqrt{2}}{4}\left(-\tanh t +\cot \theta \right)
\,,\nonumber\\
\lambda&=&-\rho 
=\frac{\sqrt{2}}{4}\left(\tanh t +\cot \theta \right).
\label{(5.7)}
\end{eqnarray}
The non-vanishing Weyl scalars are all constant:
\begin{equation}
\psi_0=\psi_4=-\frac12 \,,\qquad \psi_2=\frac{1}{6},
\label{(5.8)}
\end{equation}
from which the algebraically special nature of this metric follows easily.

By virtue of Eq. \eqref{(5.1)}, we can use the same coordinates
of Sec. II in order to define three concepts of projective
infinity. The isolated points pertaining to future timelike
infinity, spacelike infinity and past timelike infinity arise
from a projective map which engenders destruction of the 
$2$-sphere, whereas  future and past null infinity arise from
a projective map which preserves the $2$-sphere. 

On passing to new coordinates
\begin{eqnarray}
t'&=& \frac{T}{\rho}\,,\qquad
r'= \frac{R}{\rho},
\nonumber\\
\theta'&=& \frac{\Theta}{\rho},\qquad
\phi'= \frac{\Phi}{\rho},
\label{(5.9)}
\end{eqnarray}
where $\rho=1+A^0{}_1 t +A^0{}_2 r$, the metric transforms as follows
(cf. Appendix C):
\begin{widetext}
\begin{eqnarray}
g_{t't'} &=&   -\rho^4 (r'A^0{}_2 - 1)^2 
+ \rho^4 r'{}^2 \Bigr(A_{\; 1}^{0}\Bigr)^2 \cosh^2(t'\rho) 
+  \Bigr(A_{\; 1}^{0}\Bigr)^2 \rho^4\Omega'
\nonumber\\
g_{r'r'} &=&    -\rho^4 t'{}^2 \Bigr(A_{\; 2}^{0}\Bigr)^2 
+ \rho^4 (-t' A^0{}_1 + 1)^2\cosh^2(t'\rho) 
+ \Bigr(A_{\; 2}^{0}\Bigr)^2 \rho^4\Omega'
\nonumber\\
g_{\theta'\theta'} &=&\rho^2 
\nonumber\\
g_{\phi' \phi'} &=& \rho^2\sin^2(\theta'\rho) 
\nonumber\\
g_{t' r'} &=& \rho^4 (r' A^0{}_2 - 1) t'A^0{}_2 + \rho^4 r'A^0{}_1 (-t'A^0{}_1 
+ 1)\cosh^2(t'\rho) + A^0{}_1  A^0{}_2\rho^4\Omega'
\nonumber\\
g_{t' \theta'} &=& \rho^3\theta' A^0{}_1
\nonumber\\
g_{t'\phi'} &=&  \rho^3\sin^2(\theta'\rho)\phi' A^0{}_1
\nonumber\\
g_{r'\theta'} &=& \rho^3\theta' A^0{}_2
\nonumber\\
g_{r'\phi'} &=&  \rho^3\sin^2(\theta' \rho)\phi' A^0{}_2
\nonumber\\
g_{\theta'\phi'} &=& 0,
\label{(5.10)}
\end{eqnarray}
where, as before, we used the notation $\Omega'=\sin^2(\theta'\rho)\phi'{}^2 + \theta'{}^2$.

\end{widetext}

\begin{figure}
\includegraphics[scale=0.3]{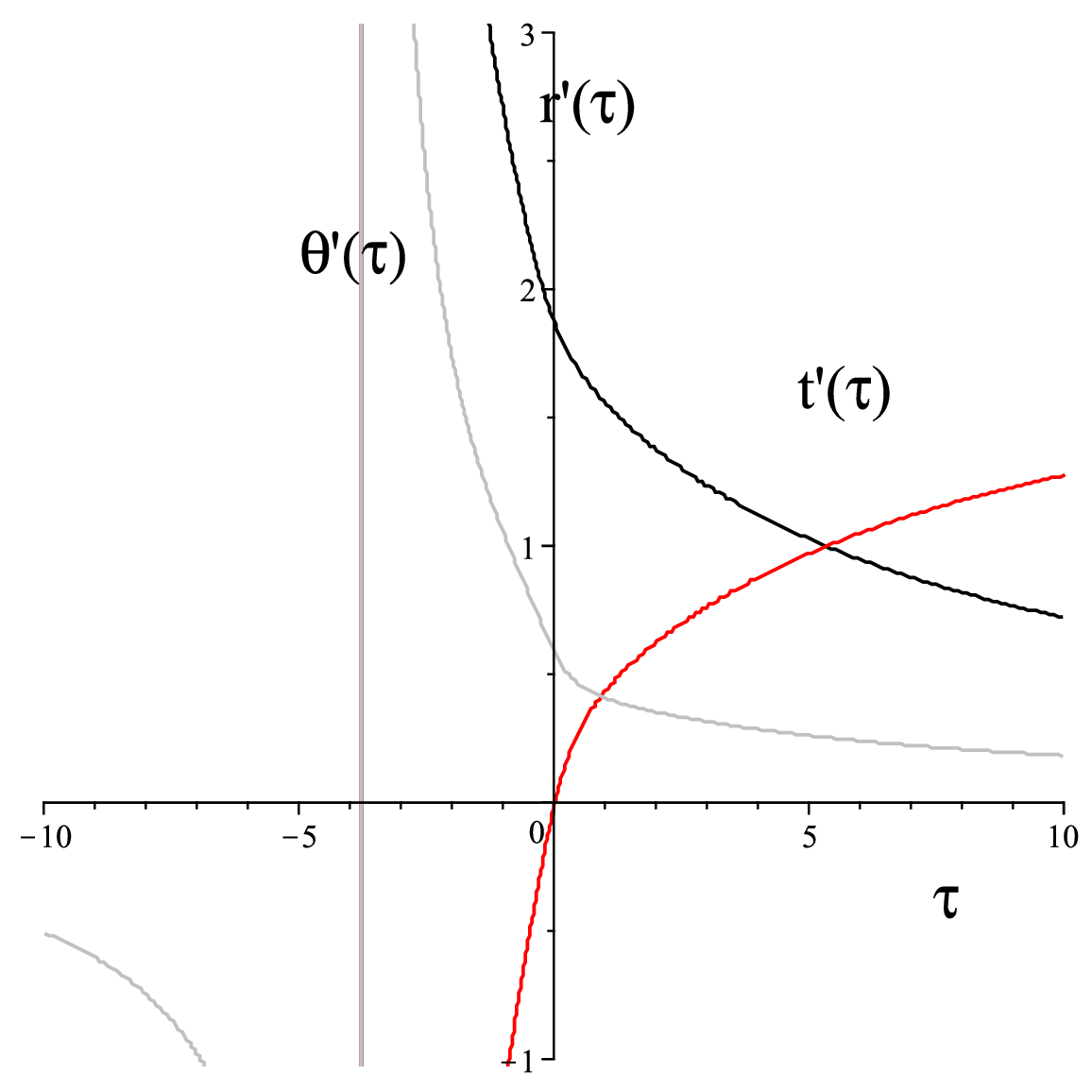}
\caption{\label{fig:xnai} Timelike equatorial plane geodesics of a 
Nariai spacetime mapped into projective coordinates.
The parameter and initial conditions choices are the same 
as in Fig. \ref{fig:5}. These values 
imply $\phi(\tau)\equiv 0$ and then it is not included in the plot.
The projective parameters chosen are $[A^0{}_1,A^0{}_2]=[\frac12,\frac13]$.
The numerical integration gives the following asymptotic values:
$r'(10^6)=1.2\cdot 10^{-5}$, $t'(10^6)=2.00$, $\theta'(10^6)=3.14\cdot 10^{-6}$.
Thus, within the numerical approximation we see that timelike geodesics  
reach our future timelike infinity $(2,0,0,0)$.  
The same qualitative behavior is obtained on considering other
choices of $A_{\; 1}^{0}$ and $A_{\; 2}^{0}$.} 
\end{figure}

\begin{figure}
\includegraphics[scale=0.3]{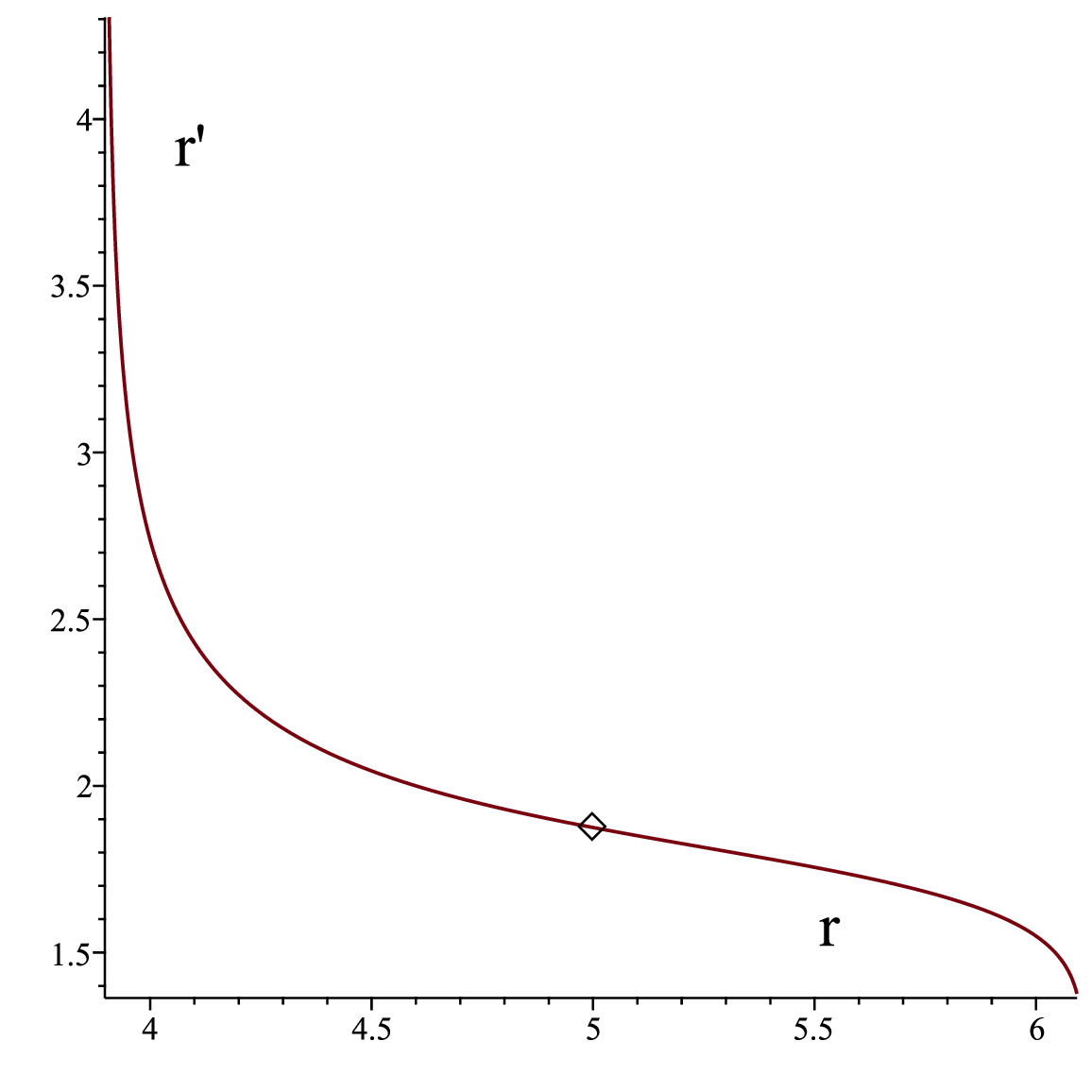}
\caption{\label{fig:x1} By using the same parameters and initial conditions of 
Fig. \ref{fig:xnai} we display the behavior of $r'$ vs $r$, with $\tau$ 
as the parametric plot parameter. The highlighted point corresponds to the value $\tau=0$.} 
\end{figure}

\section{Points at the boundary of the accessible region in a de Sitter spacetime}

Consider de Sitter spacetime, whose line element written in spherical-like coordinates
is given by
\begin{equation}
ds^2 = -N^2dt^2 +\frac{dr^2}{N^2}+r^2(d\theta^2+\sin^2\theta d\phi^2),
\end{equation}
where $N = (1 - H^2r^2)^{1/2}$ denotes the \lq\lq lapse" 
function. The spacetime region which
can be accessed is the ball of radius $r_b=1/H$ and center at the origin.
When examining any kind of motion in this spacetime the radial 
(compact) region $r\in[0,\frac{1}{H}]$ is the only one which has to be explored.
By virtue of the simplicity of the metric, with $G_{\alpha\beta}=-3H^2 g_{\alpha\beta}$, 
the geodesics are integrable as a first order system of equations
\begin{eqnarray}
\frac{dt}{d\tau} &=&  \frac{E^2}{N^2},
\nonumber\\
\frac{d\phi}{d\tau} &=& \frac{L}{r^2\sin^2\theta},
\nonumber\\
\frac{d\theta}{d\tau}&=& \frac{\epsilon_\theta}{r^2}\sqrt{C^{2}-\frac{L^2}{\sin^2\theta}},
\nonumber\\
\frac{dr}{d\tau}&=&  \frac{\epsilon_r}{r}\sqrt{E^2+(\epsilon r^2+C^2)(H^2r^2-1)},
\end{eqnarray}
where $E$ and $L$ are  Killing constants, $\epsilon=[1,0,-1]$ for timelike, 
null, spacelike orbits, respectively, $C$ is a separation constant 
for radial-polar motions and $\epsilon_r, \epsilon_\theta$ are two independent signs.
Equatorial plane motion, $\theta=\pi/2$, corresponds to 
$C=L$ and the above equations take the simpler form (Figs. 7 and 8)
\begin{eqnarray}
\frac{dt}{d\tau} &=&  \frac{E^2}{N^2},
\nonumber\\
\frac{d\phi}{d\tau} &=&  \frac{L}{r^2},
\nonumber\\
\frac{dr}{d\tau}&=&  \frac{\epsilon_r}{r}\sqrt{E^2+(\epsilon r^2+L^2)(H^2r^2-1)}.
\end{eqnarray}

\begin{figure}
\includegraphics[scale=0.3]{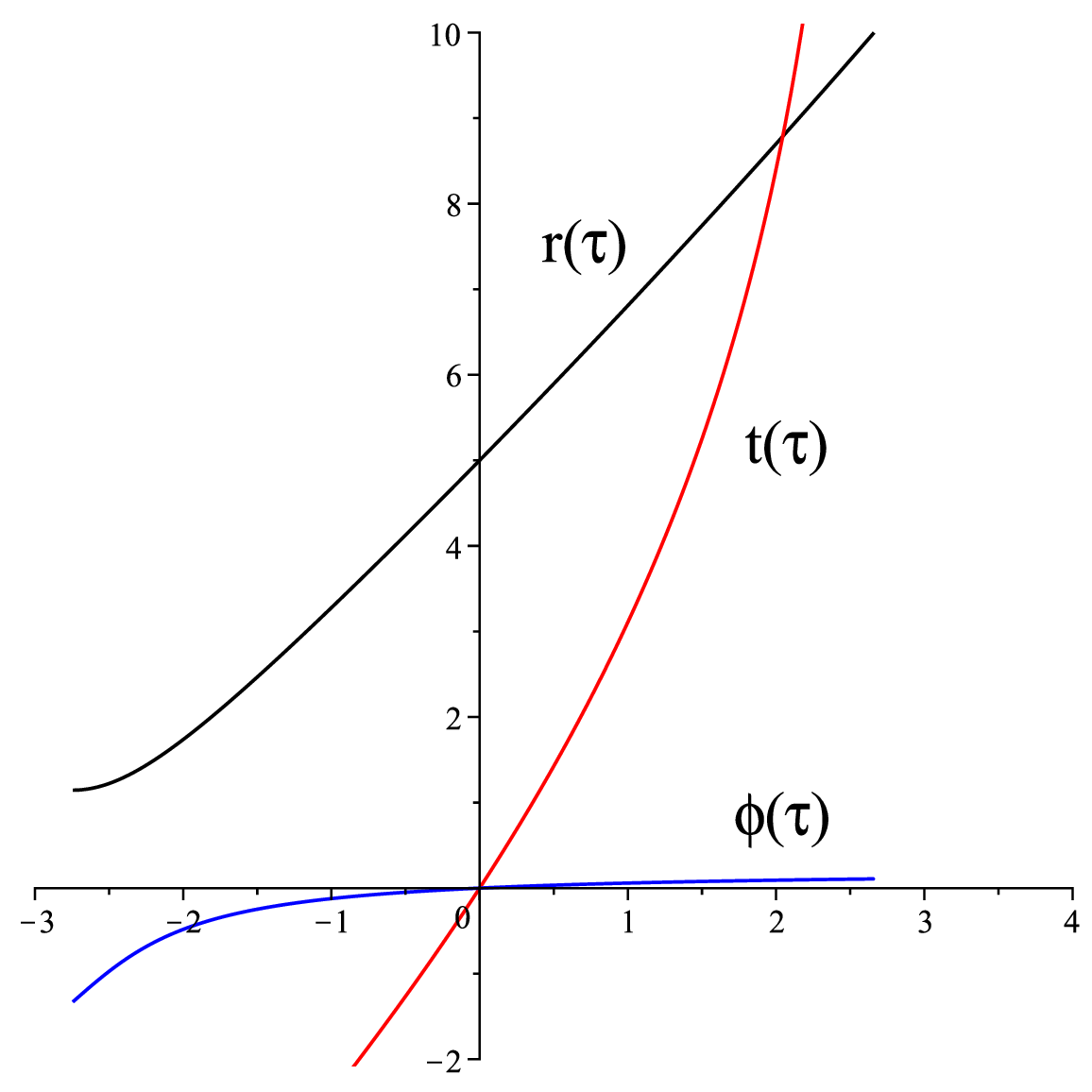}
\caption{\label{fig:dS1} Numerical integration of timelike geodesics 
in standard spherical coordinates. The parameter values are 
$H_0=\frac{1}{10}$, $E=2$, $L=2$, while initial conditions 
are $r(0)=5$, $t(0)=0$, $\phi(0)=0$.} 
\end{figure}

\begin{figure}
\includegraphics[scale=0.3]{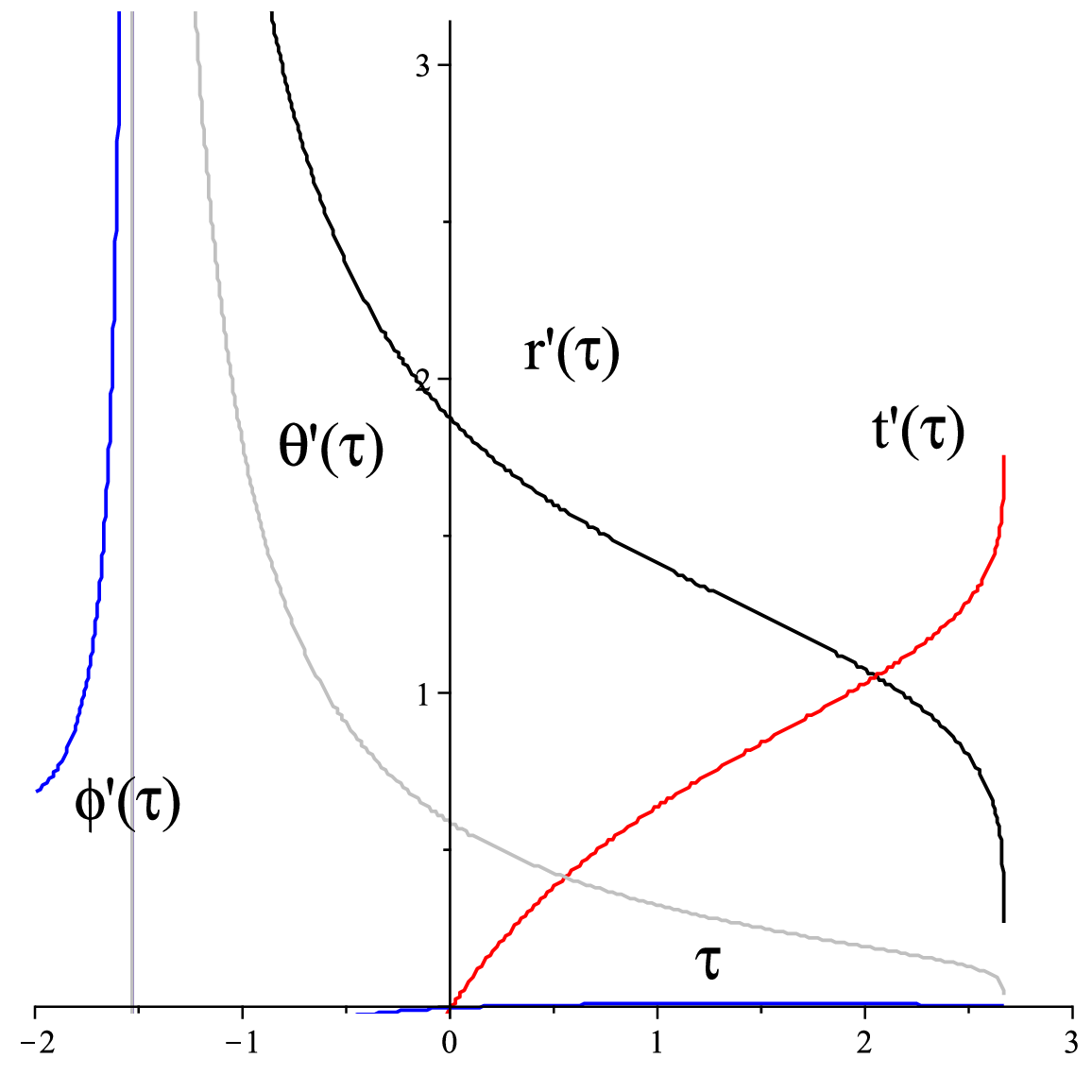}
\caption{\label{fig:dS3} Timelike equatorial plane geodesics of a 
de sitter spacetime mapped into projective coordinates.
The parameter and initial conditions choices are the same as in Fig. \ref{fig:dS1}.
The projective parameters chosen are $[A^0{}_1,A^0{}_2]=[\frac12,\frac13]$.
The maximal value of $\tau$ allowed for the numerical integration 
is $\tau_{\rm max}=2.6611$, such that
$r'(\tau_{\rm max})=0.1910$, $t'(\tau_{\rm max})=1.8344$, 
$\theta'(\tau_{\rm max})=0.0300$, $\phi'(\tau_{\rm max})=0.0021$.
}
\end{figure}

As far as we can see from Fig. $10$, timelike
geodesics reach future timelike infinity $(2,0,0,0)$, up to
the limitations resulting from numerical integration,
whereas spacelike infinity $(0,3,0,0)$ turns out to be inaccessible
(Figs. 9 and 10).

\section{Projective infinity in G\"{o}del spacetime}

A naturally occurring application of our work concerns the projective 
definition of infinity in other kinds of
spacetime. For this purpose, we have considered 
G\"{o}del's spacetime (Fig. 11). The G\"{o}del metric 
\cite{Godel,HE} is given by 
\begin{eqnarray}
\; & \; &
ds^{2}=\frac{2}{\omega^{2}} \Bigr[
-dt^{2}+dr^{2}-(\sinh^{4}r-\sinh^{2}r)d\phi^{2}
\nonumber \\
& \; & +2 \sqrt{2}\sinh^{2}r d\phi \; dt \Bigr]+dz^{2},
\label{(7.1)}
\end{eqnarray}
with coordinates
$$ 
t \in ]-\infty,\infty[, \quad
r \in [0,\infty[, \quad
\phi \in[0,2\pi],\quad
z \in ]-\infty,\infty[\,,
$$
and an observer horizon located at  $r=\ln(1+\sqrt{2})=0.8814$.
See Ref. \cite{HE} for all details. 
We provide below plots similar to those considered in the various 
examples above, showing that passing to cylindrically symmetric 
spacetimes is feasible.
\begin{figure}
\includegraphics[scale=0.3]{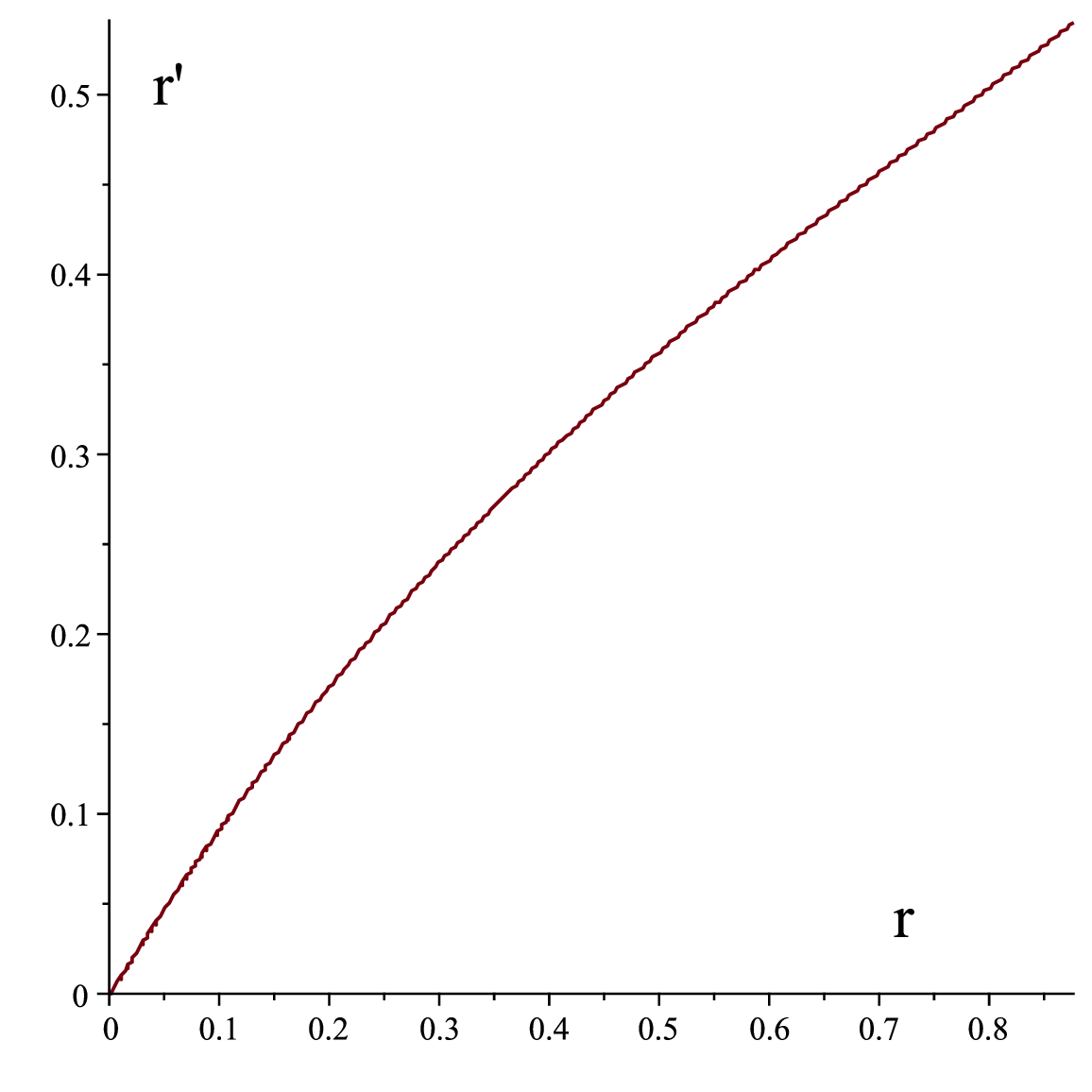}
\includegraphics[scale=0.3]{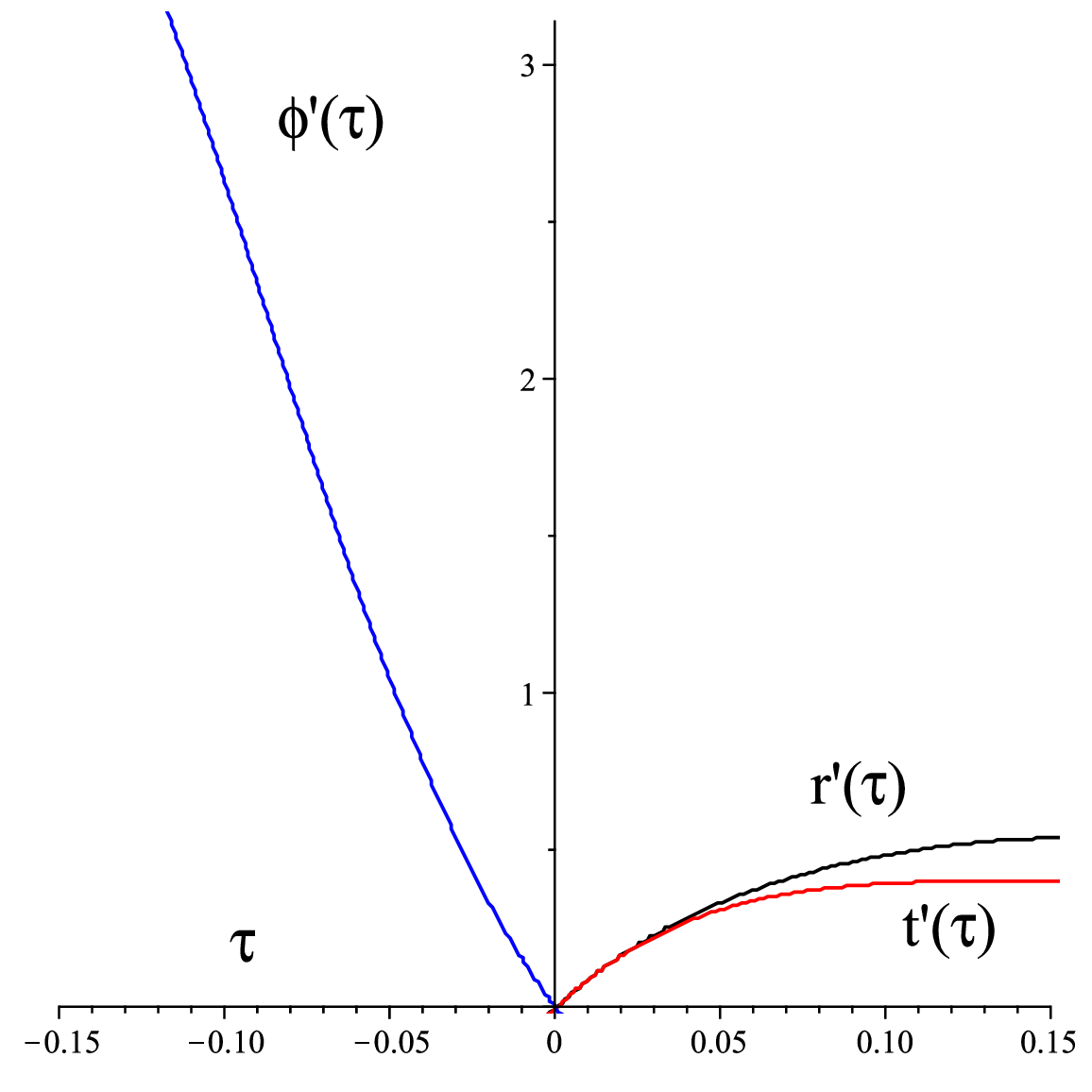}
\caption{\label{fig:dS2bis} We display the behavior of $r'$ vs 
$r$, with $\tau$ as the parametric plot parameter 
(upper panel) and the timelike geodesics on the $z=0$ plane 
(lower panel) in the G\"odel spacetime.
Spacetime and Killing parameters are chosen so that $E = 20$, 
$L = 0$, $\omega = 1$, whereas the initial conditions are taken 
at the origin: $t(0) = 0$, $r(0) = 0$, $\phi(0) = 0$. The 
projective map parameters are $A_{\; 1}^{0}=\frac12$, 
$A_{\; 2}^{0}=\frac13$, as usual.} 
\end{figure}

Here we remark instead
that, with a method analogous to the one in Sec. II, but
relying now on a matrix $A \in {\rm GL}(4,{\mathbb R})$
having the form
\begin{equation}
A=\left(\begin{matrix}
1 & A_{\; 1}^{0} & A_{\; 2}^{0} & 0  \cr
0 & 1 & 0 & 0  \cr
0 &  0 & 1 & 0  \cr
0 & 0 & 0 & 1  
\end{matrix}\right),
\label{(7.2)}
\end{equation}
we obtain the projective map
\begin{equation}
\frac{t'}{t}=\frac{r'}{r}=\frac{\phi'}{\phi}
=(1+A_{\; 1}^{0}t+A_{\; 2}^{0}r)^{-1},
\label{(7.3)}
\end{equation}
which leads in turn to three kinds of projective infinity:
\vskip 0.3cm
\noindent
(1) Projective timelike infinity, given by the point
\begin{equation}
\left(t'=\frac{1}{A_{\; 1}^{0}}={t'}_{\infty}, 
r'=0, \phi'=0 \right).
\label{(7.4)}
\end{equation}
Of course, the sign of $A_{\; 1}^{0}$ may further
distinguish among future and past timelike infinity.
\vskip 0.3cm
\noindent
(2) Projective spacelike infinity, consisting of the point
\begin{equation}
\left(t'=0,r'=\frac{1}{A_{\; 2}^{0}}={r'}_{\infty},
\phi'=0 \right).
\label{(7.5)}
\end{equation}
\vskip 0.3cm
\noindent
(3) Direct product of half-lines in the first or fourth quadrant 
with $S^1$, whose points have coordinates
\begin{equation}
\left(t'=t=-{t'}_{\infty} 
\frac{r}{{r'}_{\infty}},r'=r,\phi'=\phi \right).
\label{(7.6)}
\end{equation}
In other words, when the projective map \eqref{(7.3)} destroys the
$1$-sphere $S^1$, one deals with timelike or spacelike infinity,
whereas if the projective map preserves the $1$-sphere, one obtains
a two-dimensional manifold whose points have the coordinates
in Eq. \eqref{(7.6)}. 

\section{Concluding remarks}

When studying a certain spacetime it is of crucial importance from
both a physical and a mathematical point of view to understand its
complete causal structure. The latter contains useful information
on how the orbits of particles or photons reach special asymptotic
surfaces. Equivalently, one is interested in identifying those
hypersurfaces containing the endpoints of timelike or null orbits
(and also those of spacelike orbits, even if these hypersurfaces
might have a mainly mathematical interest). The so-called 
Carter-Penrose diagrams provide a picture of spacetime by using
compactified coordinates and the scientific community is familiar 
with this approach which, by now, does not hide special difficulties
or subtleties and is included in textbooks.

We are instead interested in a different approach and a different
point of view, closely related to abstract projective geometry
rather than directly to spacetime geometry. According to our
approach, compactified coordinates arise from the immersion of
spacetime coordinates into a projective space. In such a space
(with dimension $4+1$) we limit ourselves to natural coordinate
changes, i.e., linear transformations. In turn, these transformations
induce fractional linear maps among spacetime coordinates.

The advantage of having fractional linear coordinate maps is that
simple choices of the parameters occurring in the coordinate maps
make it possible for the new coordinates to be finite when the 
original coordinates reach infinity. A partial step in this 
direction was made in the valuable paper \cite{Philipp}, devoted
to the radial wave equation in Schwarzschild spacetime. On the
other hand, the approach therein was incomplete because a fractional
linear map was considered only for the radial coordinate, whereas
we have adopted a projective method for all coordinates of a
spherically symmetric spacetime. 
So far, the new coordinates can also be seen
as compactified coordinates as in the Carter-Penrose approach. 
In our paper we have explicitly discussed (within some special
spacetimes) the deformation induced on light cones of the given
spacetime. A different point of view consists in considering 
these coordinate transformations just as possible coordinate
changes in spacetime, ignoring their projective origin. In our
paper, following this perspective, we have studied observer
families at rest with respect to these coordinates and investigated
their geometric kinematical properties, i.e., acceleration,
vorticity and expansion.

To sum up, we believe that coordinate transformations of projective
origin shed light on the causal structure of spacetime and are
worth studying. As it usually happens in relativity, certain 
coordinate systems may allow simplifications in a problem and
complications in another. In principle, every choice is possible,
but special choices exist as well. In black hole physics for
example there is a huge variety of special observer families: 
Zero Angular Momentum observers (ZAMOS), static observer,
extremely accelerated, minimally rotating, 
Painlev\'e-Gullstrand type, and so on. We have added to such
a list what we here call projective-coordinate adapted
observers, which are useful in order to look at infinity
(looking away observers).

In more technical terms,
according to the method developed in our paper for dealing
with points at infinity, we first choose the symmetry of
interest; second, local coordinates are chosen that are
appropriate for such a case; third, such physical coordinates
are subject to fractional linear transformations. 
As far as we can see, the original contributions of our approach 
are therefore as follows.
\vskip 0.3cm
\noindent
(i) We have proved immersion of spherically symmetric
spacetime in real projective space ${\mathbb R}P_{4}$, 
and hence we have exploited fractional linear maps to
define the spacetime coordinate transformations \eqref{(2.14)}.
This makes it necessary to remove 
suitable sets\footnote{The sets of points
for which the denominator of the fractional linear
transformation vanishes. These sets reduce to lines 
in the $(r,t)$ plane when
the matrix $A$ takes the form \eqref{(2.17)}.} 
from the spacetime $(M',g')$,
and makes it possible to bring infinity down to a
finite distance.
\vskip 0.3cm
\noindent
(ii) {\it The three kinds of infinity arise from different 
implementations of one and the same concept, i.e., 
a projective map among homogeneous coordinates of
${\mathbb R}P_{4}$, which engenders a fractional linear
transformation for spacetime coordinates}. 
\vskip 0.3cm
\noindent
(iii) We can define points at infinity even in cases where
the conformal boundary does not exist, as shown in Sec. V.
\vskip 0.3cm
\noindent
(iv) We can study geodesics and observers in inhomogeneous
projective coordinates in a neat way.

Of course, a definition of infinity independent of any choice of
symmetry with the associated coordinates is impossible within
our framework. As far as we can see, no universal framework is
conceivable. Carter-Penrose diagrams do not exist in all spacetimes,
nor is null infinity always smooth in the standard 
approach \cite{Kehrberger}. 
As is clear already from Secs. II, III and IV, our projective technique
does not lead to a conformal rescaling of the original 
spacetime metric, and hence our definition of 
what we call projective counterpart of null infinity is not
the boundary of the Penrose conformal completion of a
physical spacetime (cf. the important work in Ref. \cite{AS}),
and deserves further study.
The application of our projective technique to the study
of asymptotic properties of classical and quantum gravity
is an important task in our opinion, in light of the huge
amount of work in the recent literature 
(see, for example, Refs. \cite{Andy,AE,HT,ME}).
In the future it will be also interesting to understand whether
any relation exists between the valuable work in Ref.
\cite{Shabbir} and our method.

At the risk of repeating ourselves, we find it worth stressing
once more the original contributions of our investigation: the immersion
of spherically symmetric spacetimes into real projective space has 
been built and exploited to provide a way to bring infinity down
to a finite distance, obtaining also a projective definition of
all concepts of infinity. Moreover, the possibility suggested in
Sec. II to regard null infinity as an algebraic surface offers
a novel way of looking at the asymptotic structure of spacetime.

\section*{Acknowledgements}
We are indebted to F. D'Andrea, 
M.A.H. MacCallum and J.A. Valiente Kroon for helpful 
email correspondence, and to INDAM for membership. 
G.E. is grateful to M. Taronna for several stimulating questions.  
 
\appendix

\section{Projective coordinates}

Let $r$ be the complex line supplemented by the point at
infinity, and let $\lambda$ be a coordinate on $r$. If
$S$ is a matrix of the group ${\rm GL}(2,\mathbb{C})$,
written as
\begin{equation}
S=\left(\begin{matrix}
a_{00} & a_{01} \cr
a_{10} & a_{11}
\end{matrix}\right),
\label{(A1)}
\end{equation}
the equation 
\begin{equation}
\lambda'=\frac{a_{11}\lambda+a_{10}}{a_{01}\lambda + a_{00}},
\label{(A2)}
\end{equation}
establishes a bijective correspondence, without exceptions,
between $\lambda$ and $\lambda'$, and $\lambda'$ can be
viewed as the new coordinate on $r$. 

If $\lambda,\lambda'$ are projective coordinates on the
lines $r$ and $r'$ respectively, Eq. (A2) establishes
a correspondence $\omega: r \rightarrow r'$ which is
bijective without exceptions, and is called a projective
correspondence.

\section{An alternative to the linear law \eqref{(2.10)}}

Instead of the linear law \eqref{(2.10)}, one might consider
the transformations
\begin{equation}
y_{1}'=\varphi_{1}(y_{1},...,y_{5}),...,
y_{5}'=\varphi_{5}(y_{1},...,y_{5}),
\label{(B1)}
\end{equation}
where the $\varphi_{k}$ are homogeneous polynomials, all of
the same degree $n \geq 2$. The spacetime coordinates might be
defined according to
\begin{equation}
X_{i}=\frac{y_{i}}{y_{5}}, \; \forall i=1,...,4,
\label{(B2)}
\end{equation}
with the resulting transformation law
\begin{equation}
X_{i}'=\frac{y_{i}'}{y_{5}'}
=\frac{\varphi_{i}}{\varphi_{5}}, \; \forall i=1,...,4.
\label{(B3)}
\end{equation}
The proof of the immersion of spherically symmetric spacetime
into real projective space is identical to what we have obtained
at the beginning of Sec. II, because Eqs. \eqref{(2.3)} and
\eqref{(2.4)} remain valid.

As a concrete example, let us assume that the polynomials in 
Eq. \eqref{(B1)} are homogeneous of degree $2$, so that we can write
\begin{eqnarray}
\varphi_{5}&=& B_{0}(y_{5})^{2}+...+B_{4}(y_{4})^{2}
+B_{5}y_{5}y_{1} 
\nonumber \\
&+& ...+B_{8}y_{5}y_{4}+...+B_{14}y_{3}y_{4},
\label{(B4)}
\end{eqnarray}
\begin{equation}
\varphi_{1}=\alpha_{0}(y_{5})^{2}
+\alpha_{1}(y_{1})^{2}+...+\alpha_{14}y_{3}y_{4},
\label{(B5)}
\end{equation}
\begin{equation}
\varphi_{2}=\beta_{0}(y_{5})^{2}
+\beta_{1}(y_{1})^{2}+...+\beta_{14}y_{3}y_{4},
\label{(B6)}
\end{equation}
\begin{equation}
\varphi_{3}=\gamma_{0}(y_{5})^{2}
+\gamma_{1}(y_{1})^{2}+...+\gamma_{14}y_{3}y_{4},
\label{(B7)}
\end{equation}
\begin{equation}
\varphi_{4}=\delta_{0}(y_{5})^{2}
+\delta_{1}(y_{1})^{2}+...+\delta_{14}y_{3}y_{4}.
\label{(B8)}
\end{equation}
In light of Eqs. \eqref{(B3)}-\eqref{(B8)}, we can thus write
the transformation laws (in the course of evaluating the ratios
of homogeneous polynomials of degree $2$, we divide numerator
and denominator by $(y_{5})^{2}$ and then use the definition
\eqref{(B2)})
\begin{eqnarray}
\; & \; & t'=\frac{y_{1}'}{y_{5}'}
=\frac{\varphi_{1}}{\varphi_{5}}
\nonumber \\
&=& \frac{(\alpha_{0}+\alpha_{1}t^{2}
+\alpha_{2}r^{2}+...+\alpha_{14}\theta \phi)}
{(B_{0}+B_{1}t^{2}
+B_{2}r^{2}+...+B_{14}\theta \phi)},
\label{(B9)}
\end{eqnarray}
\begin{eqnarray}
\; & \; & r'=\frac{y_{2}'}{y_{5}'}
=\frac{\varphi_{2}}{\varphi_{5}}
\nonumber \\
&=& \frac{(\beta_{0}+\beta_{1}t^{2}
+\beta_{2}r^{2}+...+\beta_{14}\theta \phi)}
{(B_{0}+B_{1}t^{2}
+B_{2}r^{2}+...+B_{14}\theta \phi)},
\label{(B10)}
\end{eqnarray}
and hence
\begin{equation}
\lim_{t \to \pm \infty}t'=\lim_{t \to \pm \infty}
\frac{\alpha_{1}t^{2}}{B_{1}t^{2}}
=\frac{\alpha_{1}}{B_{1}}=t_{\infty}',
\label{(B11)}
\end{equation}
\begin{equation}
\lim_{r \to \infty}r'=\lim_{r \to \infty}
\frac{\beta_{2}r^{2}}{B_{2}r^{2}}
=\frac{\beta_{2}}{B_{2}}=r_{\infty}'.
\label{(B12)}
\end{equation}
This implies that only two free parameters exist, i.e.,
the values of the ratios (B11) and (B12), and we can set
$\alpha_{1}=1,\beta_{2}=1$, arriving therefore at the
simpler formulas
\begin{equation}
\frac{\theta'}{\theta}
=\frac{1}{(1+B_{1}t^{2}+B_{2}r^{2})}
=\frac{\phi'}{\phi}.
\label{(B13)}
\end{equation}
At this stage, preservation of the $2$-sphere, in the sense that
$$
\theta'=\theta, \; \phi'=\phi,
$$
is obtained provided the following equation is satisfied:
\begin{equation}
B_{1}t^{2}+B_{2}r^{2}=0 \Longrightarrow 
t^{2}=-\frac{B_{2}}{B_{1}}r^{2}
=-\frac{t_{\infty}'}{r_{\infty}'}r^{2}.
\label{(B14)}
\end{equation}
Thus, in order to avoid imaginary values of $t$, $t_{\infty}'$ can
only have opposite sign with respect to $r_{\infty}'$. We find
therefore that polynomials of even degree in Eqs. \eqref{(B1)}
are less convenient than the linear law assumed in our Eq. 
\eqref{(2.10)}.

\section{All metrics with spherical symmetry}

Our Eqs. \eqref{(4.4)}-\eqref{(4.4)} can be inverted to
express the differentials $dt,dr,d\theta,d\phi$. On 
assuming the form \eqref{(2.17)} of the matrix $A$
and defining
\begin{widetext}
\begin{eqnarray}
\alpha &=& \rho (1-A_{\; 1}^{0}t')^{-1}(1+\rho
A_{\; 1}^{0} A_{\; 2}^{0}t'r'),
\nonumber \\
\beta &=& \rho^{2}A_{\; 2}^{0}t',
\nonumber \\
\gamma &=& \rho^{2}A_{\; 1}^{0}r',
\nonumber \\
\delta &=& \rho^{2}(1-A_{\; 1}^{0}t'),
\nonumber \\
{\tilde \alpha} &=& \theta'(A_{\; 1}^{0}\alpha+A_{\; 2}^{0}\gamma),
\nonumber \\
{\tilde \beta} &=& \theta'(A_{\; 1}^{0}\beta+A_{\; 2}^{0}\delta),
\nonumber \\
{\tilde \gamma} &=& \phi'(A_{\; 1}^{0}\alpha
+A_{\; 2}^{0}\gamma),
\nonumber \\
{\tilde \delta} &=& \phi'(A_{\; 1}^{0}\beta
+A_{\; 2}^{0}\delta),
\label{(C1)}
\end{eqnarray}
we find
\begin{eqnarray}
dt &=& \alpha \; dt'+\beta \; dr',
\nonumber \\
dr &=&\gamma \; dt'+\delta \; dr',
\nonumber \\
d\theta &=& {\tilde \alpha} \; dt'+{\tilde \beta} \; dr'
+\rho \; d\theta',
\nonumber \\
d\phi &=&{\tilde \gamma} \; dt'
+{\tilde \delta} \; dr'+\rho \; d\phi'.
\label{(C2)}
\end{eqnarray}
Moreover, since
\begin{equation}
(1-A_{\; 1}^{0}t')^{-1}
= \frac {\rho}{(1+\rho A_{\; 2}^{0}r')},
\label{(C3)}
\end{equation}
we obtain
\begin{equation}
A_{\; 1}^{0}\alpha+A_{\; 2}^{0}\gamma=
\frac {\rho A_{\; 1}^{0} (1+\rho A_{\; 2}^{0}r')}
{(1-A_{\; 1}^{0}t')}
= \rho^{2}A_{\; 1}^{0},
\label{(C4)}
\end{equation}
while 
\begin{equation}
A_{\; 1}^{0}\beta+A_{\; 2}^{0}\delta=
\rho^{2}A_{\; 2}^{0}.
\label{(C5)}
\end{equation}
By virtue of Eq. \eqref{(C4)}, we can solve for $\alpha$
in a more convenient form with respect to 
the first line of Eq. \eqref{(C1)}:
\begin{equation}
\alpha=\rho^{2}-\frac {A_{\; 2}^{0}} {A_{\; 1}^{0}}
\gamma=\rho^{2}(1-A_{\; 2}^{0}r').
\label{(C6)}
\end{equation}
In light of Eqs. \eqref{(C1)}-\eqref{(C6)}, we can re-express
any spherically symmetric metric
\begin{equation}
g = g_{tt} dt \otimes dt + g_{rr} dr \otimes dr
+ g_{\theta \theta} d\theta \otimes d\theta
+g_{\phi \phi} d\phi \otimes d\phi,
\label{(C7)}
\end{equation}
in nondiagonal form, with components
\begin{eqnarray}
g_{t't'}&=& \rho^{4}\Bigr[g_{tt}(1-A_{\; 2}^{0}r')^{2}
+g_{rr}(A_{\; 1}^{0})^{2}{r'}^{2}\Bigr]
+ \rho^{4}(A_{\; 1}^{0})^{2}
(g_{\theta \theta}{\theta'}^{2}
+g_{\phi \phi}{\phi'}^{2}),
\nonumber \\
g_{t'r'}&=& \rho^{4}\Bigr[g_{tt}A_{\; 2}^{0}
(1-A_{\; 2}^{0}r')t'+g_{rr}A_{\; 1}^{0}r'
(1-A_{\; 1}^{0}t')\Bigr]
+ \rho^{4}A_{\; 1}^{0} A_{\; 2}^{0}
(g_{\theta \theta}{\theta'}^{2}+g_{\phi \phi}{\phi'}^{2}),
\nonumber \\
g_{r'r'}&=& \rho^{4}\Bigr[g_{tt}
(A_{\; 2}^{0})^{2}{t'}^{2}
+g_{rr}(1-A_{\; 1}^{0}t')^{2}\Bigr]
+ \rho^{4}(A_{\; 2}^{0})^{2}
(g_{\theta \theta}{\theta'}^{2}
+g_{\phi \phi}{\phi'}^{2}),
\nonumber \\
g_{t'\theta'}&=&g_{\theta \theta} \theta' \rho^{3} A_{\; 1}^{0},
\nonumber \\
g_{r'\theta'}&=&g_{\theta \theta} \theta' \rho^{3} A_{\; 2}^{0},
\nonumber \\
g_{t'\phi'}&=&g_{\phi \phi} \phi' \rho^{3} A_{\; 1}^{0},
\nonumber \\
g_{r'\phi'}&=&g_{\phi \phi} \phi' \rho^{3} A_{\; 2}^{0},
\nonumber \\
g_{\theta' \theta'}&=&g_{\theta \theta} \rho^{2},
\nonumber \\
g_{\phi' \phi'}&=&g_{\phi \phi} \rho^{2},
\nonumber \\
g_{\theta' \phi'}&=&0.
\label{(C8)}
\end{eqnarray}
\end{widetext}

\end{document}